\documentclass[prx,aps,floatfix,longbibliography,superscriptaddress,twocolumn,final,notitlepage]{revtex4-2}
\RequirePackage{amsmath}
\RequirePackage{hyperref} 
\RequirePackage{amssymb}
\RequirePackage{graphicx}
\RequirePackage{bbm}
\RequirePackage{xcolor}
\usepackage{lipsum}
\usepackage{comment}

\usepackage{pgf}

\setcitestyle{round}

\usepackage{mathtools}
\usepackage{caption}
\usepackage{subcaption}
\usepackage{amsthm}

\newcommand{\be}{\begin{equation}}
\newcommand{\ee}{\end{equation}}
\newcommand{\bea}{\begin{eqnarray}}
\newcommand{\eea}{\end{eqnarray}}

\newcommand{\ccup}[1]{\left\{#1\right\}}

\newcommand{\bup}[1]{\left(#1\right)}
\newcommand{\rup}[1]{\left[#1\right]}
\newcommand{\pois}{\textrm{Pois}}
\newcommand{\dmax}{D}
\newcommand{\Hyg}{\mathcal{H}}  
\newcommand{\V}{V}  
\newcommand{\w}{A}  
\newcommand{\hyespace}{\Omega}  
\newcommand{\normconst}{\kappa}  
\newcommand{\degseq}{d}
\newcommand{\dimseq}{k}

\DeclareMathOperator*{\argmax}{arg\,max}

\newcommand{\algoname}{Hy-MMSBM}
\newcommand{\repolink}{\href{http://github.com/nickruggeri/Hy-MMSBM}{github.com/nickruggeri/Hy-MMSBM}}

\newtheorem{theorem}{Theorem}

\newtheorem*{definition*}{Definition}

\usepackage{amsmath,fixmath}
\usepackage{setspace}

\usepackage[noline, linesnumbered]{algorithm2e}
\RestyleAlgo{ruled}
\SetKwComment{Comment}{> }{}
\SetKwInput{KwInput}{Input}
\SetKwInput{KwYield}{Yield}

\usepackage{float}
\usepackage[nameinlink,capitalize]{cleveref} 

\usepackage{booktabs}
\usepackage{adjustbox}
\definecolor{shadecolor}{gray}{0.9}
\usepackage{tabu}

\usepackage[normalem]{ulem} 
\usepackage{soul}

\usepackage{enumitem,amssymb}
\newlist{todolist}{itemize}{2}
\setlist[todolist]{label=$\square$}
\newlist{todolist_done}{itemize}{2}
\setlist[todolist_done]{label=$\blacksquare$}
\usepackage{pifont}
\begin{document}

\title[
{A framework to generate hypergraphs with community structure} 
    ]{
{A framework to generate hypergraphs with community structure}
}

\author{Nicol\`o Ruggeri}
    \email{nicolo.ruggeri@tuebingen.mpg.de}
	\affiliation{Max Planck Institute for Intelligent Systems, Cyber Valley, 72076 Tübingen, Germany}
	\affiliation{Department of Computer Science,  ETH,  8004 Z\"urich, Switzerland}

\author{Federico Battiston}
	\email{battistonf@ceu.edu}
	\affiliation{Department of Network and Data Science, Central European University, 1100 Vienna, Austria}

\author{Caterina De Bacco}
	\email{caterina.debacco@tuebingen.mpg.de}
	\affiliation{Max Planck Institute for Intelligent Systems, Cyber Valley, 72076 Tübingen, Germany}

\begin{abstract}
In recent years hypergraphs have emerged as a powerful tool to study systems with multi-body interactions which cannot be trivially reduced to pairs. 
While highly structured methods to generate synthetic data have proved fundamental for the standardized evaluation of algorithms and the statistical study of real-world networked data, these are scarcely available in the context of hypergraphs.
Here we propose a flexible and efficient framework for the generation of hypergraphs with many nodes and large hyperedges, which allows specifying general community structures and tune different local statistics. We illustrate how to use our model to sample synthetic data with desired features (assortative or disassortative communities, mixed or hard community assignments, etc.), analyze community detection algorithms, and generate hypergraphs structurally similar to real-world data. Overcoming previous limitations on the generation of synthetic hypergraphs, our work constitutes a substantial advancement in the statistical modeling of higher-order systems.
\end{abstract}

\maketitle

\section{Introduction}
\label{sec:intro}
Over the last decades, networks have emerged as a fundamental tool to describe complex relational data in nature, society and technology~\cite{boccaletti2006complex}. Indeed, most real-world systems are nowadays known to be characterized by a highly non-trivial organization, which includes triadic closure and high clustering~\cite{watts1998collective}, low diameter and an efficient communication structure~\cite{latora2001efficient}, and unequal degree distributions~\cite{barabasi1999emergence}. Noticeably, many systems reveal the existence of modules or communities, where nodes are naturally clustered in different groups based on their patterns of connections~\cite{fortunato2010}. Identifying communities is an important task that allows performing various downstream analysis on networks, describing the roles of nodes and, generally, providing a low dimensional representation of possibly large systems. Since the seminal papers by Newman et al.~\cite{newman2004finding} and Lanchichenetti et al.~\cite{lancichinetti2008benchmark}, the problem of generating synthetic data for highly structured graphs with prescribed features has attracted enormous interest in the community. 
On the one hand, these models have led to tremendous improvements in evaluating which community detection algorithms perform best at a given task~\cite{fortunato2016community}. 
On the other hand, they have allowed the reliable generation of large synthetic data samples, useful to analyze non-trivial statistics from single instances of real networks and systematically investigate the impact of mesoscale structure on dynamical processes on graphs \cite{arenas2006synchronization, nematzadeh2014optimal}. 
This methodology has been applied to different domains, including studies on polarization on social media \cite{coscia2022minimizing}, percolation thresholds in brain networks \cite{bordier2017graph}, and structural and covariate information \cite{lusher2013exponential,hobson2021guide}.

Despite their success, recent evidence suggests that graphs can only provide a limited description of reality, as links are inherently limited to describe pairwise interactions~\cite{battiston2020networks, torres2021why, battiston2021physics, battiston2022higher}.
By contrast, non-dyadic higher-order interactions have been observed across different domains, including the human brain~\cite{petri2014homological,giusti2016two, santoro2022unveiling}, collaboration networks~\cite{patania2017shape}, species interactions~\cite{grilli2017higher}, cellular networks~\cite{klamt2009hypergraphs}, drug recombination~\cite{zimmer2016prediction}, and face-to-face human~\cite{cencetti2021temporal} and animal~\cite{musciotto2022beyond} interactions. Interestingly, such higher-order interactions naturally lead to the emergence of new collective phenomena in synchronization~\cite{bick2016chaos,skardal2020higher,millan2020explosive, lucas2020multiorder,gambuzza2021stability} and contagion~\cite{iacopini2019simplicial,chowdhary2021simplicial,neuhauser2020multibody} dynamics, diffusive process~\cite{schaub2020random,carletti2020random} and evolutionary games~\cite{alvarez2021evolutionary, civilini2021evolutionary}. 
Hypergraphs~\cite{berge1973graphs}, where hyperedges encode interactions among an arbitrary number of system units, are a natural framework to describe relational data beyond the pair~\cite{battiston2020networks}. 
In the last few years many tools have been developed to characterize the higher-order organization of real-world hypergraphs, including new centrality measures~\cite{benson2019three, tudisco2021node}, higher-order clustering~\cite{benson2018simplicial} and motif analysis~\cite{lotito2022higher}, hypergraph backboning~\cite{musciotto2021detecting}, hyperedge prediction~\cite{contisciani2022principled}, methods to infer higher-order interactions from low-order data~\cite{young2021hypergraph}. 
In particular, several tools to extract higher-order communities have been proposed, either based on flow distribution~\cite{carletti2021random, eriksson2021choosing} or statistical inference frameworks~\cite{contisciani2022principled, chodrow2021generative}.

Nevertheless, how to generate structured hypergraphs is still an open problem. The few currently available models mainly focus on ``unstructured'' higher-order generalizations of the configuration~\cite{courtney2016generalized,young2017construction,chodrow2020configuration} and the Erdos-Renyi model~\cite{barthelemy2022class}, or on growth models for hypergraphs~\cite{kovalenko2021growing,millan2021local,krapivsky2022random}. A different perspective is that of relational hyperevent models~\cite{lerner2019rem}, which specify event rates based on hyperedge statistics for hyperedges to exist, similarly to what exponential random graphs do for networks~\cite{robins2007introduction,park2004statistical}. All these approaches, however, do not account for community structure, hence are of limited usage when it comes to reproducing the complex mesoscale organization of real-world higher-order systems.
Recent works introduced latent variables models to infer community structure in hypergraphs~\cite{mulder2021latent, contisciani2022principled,chodrow2021generative}, however they do not explain how to sample from the generative model. Indeed, while sampling and inference are often studied jointly in standard networks, these two tasks present distinct computational and theoretical challenges in the case of hypergraphs. 
 
In this work, we provide a principled and general framework to sample hypergraphs. 
In particular, our method allows flexible sampling of higher-order networks with prescribed microscale and mesoscale features, controlling the distribution of node degrees and hyperedge sizes, as well as specifying arbitrary community structure (e.g. hard vs overlapping membership, assortative vs disassortative, etc.). The method is highly efficient, and scales well with the number of nodes, hyperedges, as well as hyperedge size, making it suitable for the analysis of real-world systems. In the following, we first introduce  our generative model and sampling strategy. Then, we extensively characterize the hypergraphs obtained by investigating the phase space associated with the different structural parameters.
Finally, we show how to utilize our method to analyze the structural and statistical properties of real-world data.

\section{Generative model}
\label{sec:generative model}
\begin{figure}[t]
\centering
\includegraphics[width=0.38\textwidth]{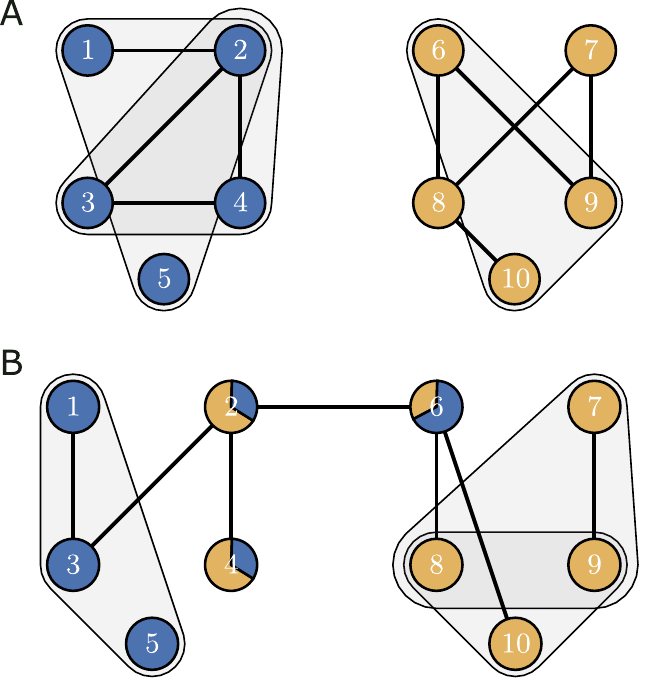} 
\caption{
    \textbf{Sampling hypergraphs with community structure.} A pictorial representation of two small hypergraphs with $N=10$ nodes, $K=2$ communities, and (A) hard or (B) overlapping membership assignment. Every node's membership assignment $u_i = (u_{i1}, u_{i2})$ is represented as a pie chart. Single colored nodes have hard assignments, mixed charts represent overlapping assignments. Due to the likelihood in \cref{eq: lambda}, nodes with overlapping assignments are more likely to belong to between-community interactions. 
    }
\label{fig:teaser}
\end{figure}
We consider hypergraphs $\Hyg(\V, E)$ consisting of $N$ nodes $V=\{1, \ldots, N\}$ and a hyperedge set $E$, where each hyperedge $e \in E$ describes an interaction among an arbitrary set of unique nodes, i.e. $e \subseteq V$, and $|e|$ is the hyperedge size.
The degree of a node $i$, i.e. the number of hyperedges it belongs to, is denoted as $\degseq_i$.
Similarly, we define the degree sequence $\degseq=\ccup{\degseq_1,\dots,\degseq_N}$ as the vector of node degrees and the size sequence $\dimseq=\ccup{k_1,\dots,k_D}$ as the count of hyperedges per hyperedge size \cite{chodrow2020configuration}. 
We consider hyperedges of arbitrary sizes, up to a maximum of $\dmax \le N$, and denote the space of all possible such hyperedges with $\hyespace$.  
We assume positive and discrete hyperedge weights, encoded using a vector $\w \in \mathbb{N}^{|\hyespace|}$, so that $E =\{e \in \hyespace : A_e > 0 \}$.

Our sampling approach introduces a flexible way to generate highly structured weighted hypergraphs with mesoscale structure, where hyperedges are generated probabilistically and nodes belong to $K$ communities.
Specifically, each node $i \in \V$ is assigned a $K$-dimensional membership vector $u_i$, where we allow $u_{ik}\geq 0$ for the general case of soft membership, where nodes can belong to multiple communities.
The particular case of hard membership assignment, where a node can only belong to one community, is recovered by setting only one non-zero entry for $u_i$. In \cref{fig:teaser} we illustrate these two cases by showing two small hypergraphs with hard or overlapping community structure. The non-negative symmetric $K\times K$-dimensional \textit{affinity} matrix $w$ regulates the interactions between communities. Classic patterns are assortative affinity matrices, with dominant diagonal signaling stronger inter-community interactions, and disassortative ones, where the out-diagonal terms have higher magnitude.  
For any given hypergraph, we define the following likelihood function:
\begin{align}
p(\Hyg ; w,  u)
	&= \prod_{e \in \Omega} p(A_e ; u,w) \nonumber \\
	&= \prod_{e \in \Omega} \pois\left(A_e ; \frac{\lambda_e}{\normconst_{|e|}} \right) \, , \label{eq: factorized prob}
\end{align}
where  
\begin{equation}
\label{eq: lambda}
\lambda_e := \sum_{i < j \in e} u_i^T \, w \, u_j \,=\sum_{i < j \in e} \sum_{k,q =1}^K \,u_{ik} w_{kq} u_{jq}.
\end{equation}
This parameterization allows generating hypergraphs under different scenarios, e.g. with assortative or disassortative community structures, and is reminiscent of those used in probabilistic models for pairwise networks \cite{debacco2017community,schein2015bayesian} and in variants of non-negative tensor factorization as used in the machine learning community \cite{kolda2009tensor,lee1999learning} when $\dmax = 2$. In addition, restricting our model to $\dmax=2$ and $\normconst_2=1$ recovers the canonical Poisson stochastic block model \cite{karrer2011stochastic}.
The parameter $\normconst_{|e|}$ is a normalization factor and is a function of the size $|e|$ of the hyperedge $e$ only (i.e. it only depends on the size of the interaction, and not on the nodes involved in it). 
These constants regulate the expected statistics of the model, such as expected degree and hyperedge size distribution. In general, any choice of $\normconst_d > 0$ yields a well-defined probabilistic model. We illustrate sensible values for $\normconst_d$ in \cref{sec supp: kappa}.

Alternative generative models for hypergraphs have been recently proposed. In particular, the works of Chodrow et al.~\cite{chodrow2021generative} and Contisciani et al.~\cite{contisciani2022principled} can be more closely compared to the model in \cref{eq: factorized prob}, since they are both based on factorized Poisson likelihoods based on communities. The former work assumes sufficient statistics only evaluated on hard community assignments and we are not aware of any computationally efficient sampling procedure from the relative generative process. 
The model of Contisciani et al.~\cite{contisciani2022principled}, instead, bears closer resemblance to the one proposed in this paper. 
The main difference lies in the specific form of the Poisson means, which, for every hyperedge $e$, are based on a product of $|e|$ terms, as opposed to the bilinear form in \cref{eq: factorized prob}. 
Despite the similar generative process, the tools utilized in this work cannot be straightforwardly applied to that model, as closed-form statistics and approximate Central Limit Theorem results cannot be derived in the same manner.\\
More generally, the primary goal of the aforementioned models is to infer hypergraph structure, leaving the problem of sampling unsolved. While our model is also well suited to efficiently infer hypergraph structure, as we illustrate in Ruggeri et al.~\cite{ruggeri2022inference}, the primary objective of this work is to demonstrate how we can effectively sample from its probability distribution. This key model's capability makes it possible to generate highly structured synthetic data with higher-order interactions. This is a key advancement for practitioners handling hypergraph data and follows influential work on such a topic for pairwise networks \cite{newman2004finding,lancichinetti2008benchmark}.

\section{Sampling hypergraphs}
\label{sec: sampling}
We now propose an efficient way to sample hypergraphs from the generative model defined in \cref{eq: factorized prob}.
Such a task is far from being straightforward.
To see why, let us consider a pairwise network model, where the configuration space is of size $|\hyespace| = N^2$, and compare it with our higher-order problem. In the former case, generation is feasible by simply exploring every single edge separately and sampling from the relative Poisson distribution. In the latter case, however, the rapid growth of the $\hyespace$ space renders both naive sampling techniques and Monte Carlo algorithms inapplicable. Here, we propose a solution to this challenge using approximate sampling. 
In the following, we focus on the intuition behind our method and illustrate relevant usage example. For a more technical description we defer to \cref{sec supp: sampling}.

\subsection{Sampling algorithm}
\label{sec: sampling algo}
Our sampling procedure follows three consecutive steps:

\paragraph{Sampling node degrees and hyperedge sizes.} The first sampling step consists of approximately sampling the $\degseq$ and $\dimseq$ vectors for a given choice of community memberships $u$ and affinity matrix $w$. Then, we use these two quantities to draw a first proposal of a binary hypergraph defined by the array $\w^b \in \{0, 1\}^{|\hyespace|}$.
More in detail, we first approximate $p(\degseq, \dimseq; u, w) \approx p(\degseq; u, w) \, p(\dimseq; u, w)$ and then use the Central Limit Theorem (CLT) to sample from $p(\degseq; w, u)$ and $p(\dimseq; w, u)$ separately. We note that these are the only approximations needed in the whole sampling routine. We elaborate more on their validity in \cref{sec supp: merging sequences}.
After sampling the $d, k$ sequences, we combine them into a first binary hypergraph configuration (i.e. a list of hyperedges) to be passed in input to the next sampling step. Intuitively, we incrementally build a hyperedge list until exhaustion of both sequences, starting by first taking the nodes with highest degrees. If the two sequences are not compatible, i.e. it does not exist a hypergraph that satisfies both, one can choose which of the two sequences to preserve during the hyperedge list construction. Such sequence will be exactly replicated, while the other will be modified to construct the first list proposal. Notice that the recombination problem has connections with the Havel-Hakimi algorithm~\cite{hakimi1962realizability} and the Erd\"os-Gallai Theorem~\cite{choudum1986simple}. Hence, the algorithm we propose for this task is a technical novelty of independent interest. We explain the algorithm in detail and present a pseudocode for it in \cref{sec supp: merging sequences}.

\paragraph{Sampling hyperedges.} 
In this second step, we sample the binary hyperedges $\w_e^b$, conditioned on $\degseq$ and $\dimseq$, using a Markov Chain Monte Carlo (MCMC) routine. This works by continuously mixing the hyperedges starting from the initial proposal $\w^b$ obtained at step \textit{a}.
The main tool utilized here is the reshuffling operator introduced in Chodrow et al.~\cite{chodrow2020configuration}: given two hyperedges $e_1,e_2$,  reshuffle the nodes not belonging to the intersection $e_1 \cap e_2$ to obtain two new hyperedges $e_1',e_2'$.  Then, accept or reject the new proposal according to the Metropolis-Hastings algorithm \cite{hastings1970monte}, whose acceptance rates depend on the Poisson means $\lambda_{e_1} / \kappa_{e_1}, \lambda_{e_2} / \kappa_{e_2}$ and consequently on the $u, w$ parameters.
Due to the properties of the reshuffling operator the new hyperedges  $e_1',e_2'$ have same sizes as $e_1,e_2$, hence the sequences $\degseq$ and $\dimseq$ are preserved. Intuitively, the Markov chain achieves good mixing owing to conditioning on $(\degseq,\dimseq)$, which restricts the space of the possible configurations.  

\paragraph{Sampling hyperedge weights.} In the third and final step, we sample the weights $\w_e$ from $p\bup{\w_e | \w_e^b=1}$. This conditional distribution is a zero-truncated Poisson with mean $\lambda_e / \normconst_{|e|}$. A related efficient sampling procedure based on inverse transform sampling is proposed in~\cref{sec supp: sampling step 3}.

Altogether, the three sampling steps described above correspond to the following probabilistic decomposition:
\begin{equation}
p\bup{A ;u, w} = p\bup{A | A^b; u, w}\,p\bup{A^b | \degseq, \dimseq; u, w}\,p\bup{\degseq, \dimseq; u, w} \, . \label{eq: sampling factorization}
\end{equation}

We provide the pseudocode of the sampling procedure in~\cref{alg:sampling} and provide an open-source implementation at \repolink.

\begin{algorithm}[H]
\caption{
    Sampling algorithm. \\
    \textit{a}: Lines~1-3; \textit{b}: Lines~4-10; \textit{c}: Line~11. 
}
\label{alg:sampling}
\KwInput{Number of communities $K$, memberships $u$, affinity $w$, MCMC burn-in steps $n_b$ and intermediate steps $n_i$, number of samples $S$.} 
\KwResult{$\ccup{A^{(s)}}_{s=1,\dots,S}$}
Sample binary degree sequence $\degseq \sim p\bup{d; u, w}$  \\
Sample size sequence $\dimseq \sim p\bup{k; u, w}$ \\
Create first proposal $\w^b$ from $d, k$  \label{algoline: create first proposal} \\
\For{$i=1, \ldots, n_b$  \label{algoline: MCMC burnin}
}{  
  	$A^b \gets $ reshuffle$(A^b)$, accept according to Metropolis-Hastings, depending on $(u, w)$
  }
\For{$s=1, \ldots, S$}{
  \For{$i=1, \ldots, n_i$ }{
  	$A^b \gets $ reshuffle$(A^b)$, accept according to Metropolis-Hastings, depending on $(u, w)$
  }
  sample $A^{(s)} \sim p\bup{A | A^b; u, w}$ \\
  yield $A^{(s)}$
}
\end{algorithm}\vspace{-2mm}

\subsection{Additional user input}
\label{sec: additional user input}
The sampling procedure described above only requires the community assignments $u$ and affinity matrix $w$ as generative parameters. 
However, a practitioner may desire to generate hypergraphs with specific features, such as a given degree or hyperedge size sequence.
Our model allows doing so naturally, either by providing such statistics as additional input or by tuning the generative parameters prior to sampling.
More precisely, one can skip the initial step and simply fix $\degseq$ or $\dimseq$ (or both) as input instead of sampling them. 
As explained in \cref{sec: sampling algo}, these quantities are guaranteed to be preserved in the sampled hypergraphs. Algorithmically, this corresponds to starting directly from line \cref{algoline: create first proposal} in \cref{alg:sampling}. 

In some cases, one might be interested in replicating the $\degseq^{data}, \dimseq^{data}$ sequences observed in a real hypergraph dataset. 
In such a simplified scenario, one can condition on the (binarized) hyperedges of the data, and proceed by directly mixing them via the MCMC procedure in the second sampling step. Since the hyperedges define the degree and size sequences, these will be preserved and identical to those of the real data, while the samples will come from the model's probability distribution. As per \cref{eq: sampling factorization}, the MCMC procedure will yield samples from $p(A^b | \degseq^{data}, \dimseq^{data}; u, w)$.
Notice that, in general, conditioning on any given sequence $d$ or $k$ might yield samples $A$ outside the high-density areas of the distribution. This is a desirable feature, as it allows the user to further specify constraints and sample hypergraphs that would otherwise be far from the typical samples obtained without conditioning \cite{erdosa2016second}.\\
Finally, with our model we can obtain closed-form expressions for relevant hypergraph properties in terms of $u$ and $w$, e.g. the expected degree of nodes, as shown in \cref{sec supp: expected stats}. 
This means that, by tuning the $u, w$ parameters, such properties can be specified prior to sampling. We illustrate some examples of this procedure in \cref{sec: synthetic data}.

\begin{figure*}[t]
\centering
    \includegraphics[width=1.0\textwidth]{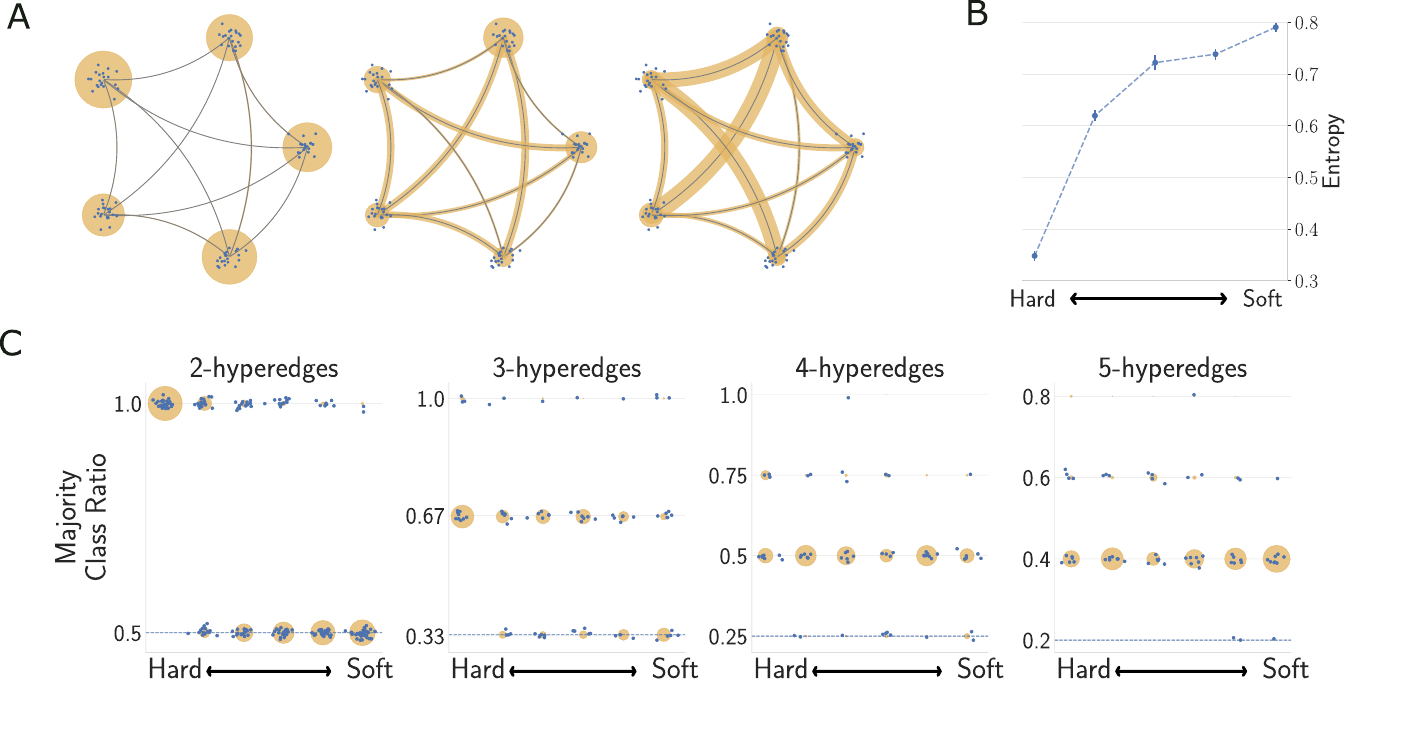}
\caption{
\textbf{Sampling hypergraphs with hard and soft community assignment}.
(A) We sample hypergraphs from a model with $K=5$ equally-sized communities, an assortative affinity matrix $w$, and different node community memberships $u$ (from hard to soft). The five shaded yellow circles represent different communities, the thicknesses of the edges and circles are proportional to the interaction strength between and within communities. (B) The entropy of community memberships grows as increasingly overlapping configurations are considered.
(C) We show the maximum assignment ratio (the relative number of nodes belonging to the majority class for each hyperedge) across hyperedge sizes. Orange circles are proportional to the amount of hyperedges with a given maximum assignment ratio. 
}
\label{fig: community assignment plots} 
\end{figure*}

\section{Synthetic Data}
\label{sec: synthetic data}
In this section, we illustrate how the generative parameters $u$ and $w$ can be tuned to sample hypergraphs with desired structures at a micro (node and hyperedge) and mesoscale (community structure and hypergraph-level statistics) level. We release ready-to-use examples of these synthetic datasets along with the open-source implementation.

\subsection{Community assignment}
We begin by showing how varying the overlap in the membership assignments $u$ leads to different intra and inter-commmunity structure. In \cref{fig: community assignment plots} we tune the assignments from hard ($u_i$ has only one non-zero entry), to soft  ($u_i>0$ for multiple entries), and highlight the strength of the interactions between and within communities by varying the thickness of edges and circles.
As memberships vary from hard to soft (left to right), edges become thicker and circles smaller, as inter(intra)-community interactions increase (decrease). Quantitatively, we compute the entropy $-\sum_{k=1}^K r_k \log r_k$, where $r_k$ is the ratio of nodes belonging to community $k$. In mixed-membership settings, one can extract a proxy for a hard assignment for node $i$ by selecting the $k = \argmax_k u_{ik}$; we use this to compute $r_k$. Lower entropy denotes hyperedges whose nodes mostly belong to the same communities, higher values denote hyperedges with nodes distributed across different communities. In \cref{fig: community assignment plots}B we show how the entropy of the community distribution grows as we sample from increasingly overlapping models. We also study the partition in communities of nodes belonging to hyperedges of different sizes. For each hyperedge we compute the ratio of nodes that belong to the majority class. For example, in a hyperedge of size 5 with two nodes in class 1 and three in class 2, the majority class is 2, yielding a majority class ratio of $3/5$. \cref{fig: community assignment plots}C shows how this ratio decreases going from hard to soft memberships, illustrating the heterogeneity of the nodes' communities across hyperedges of different sizes. 

\subsection{Affinity matrix and heterogenous \\community size}
\begin{figure*}[t]
\centering
\includegraphics[width=\textwidth]{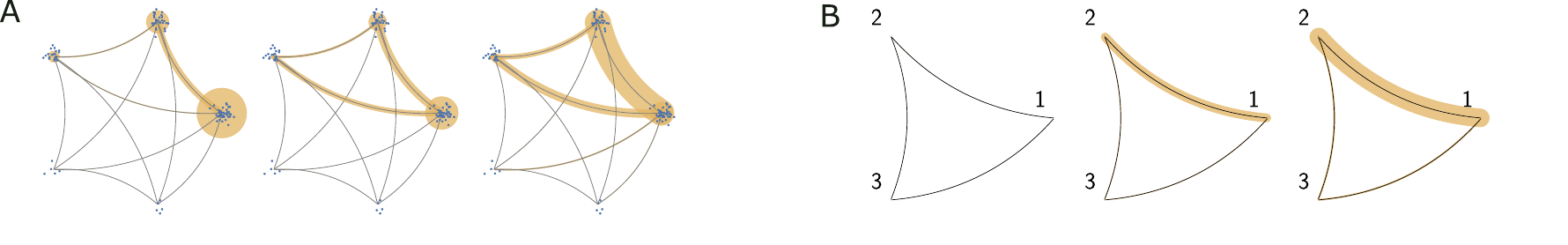}
\caption{   
    \textbf{Sampling hypergraphs with assortative and disassortative affinity and heterogeneous community size}. 
    (A) We sample hypergraphs with five communities of different sizes and hard membership assignments. We vary the affinity matrix $w$ from assortative (left, diagonal) to disassortative (right, uniform matrix filled with ones). Shaded yellow circles represent the communities, the thicknesses of the edges and circles are proportional to the interaction strength between and within communities. (B) We vary the affinity $w$ from diagonal (left) and increase its entries $w_{12}, w_{21}$ (right) for $K=3$ equally-sized communities. Nodes represent communities and the thickness of the edges and circles is proportional to the strength of the interactions between and within communities.
}
\label{fig: affinity and community size}
\end{figure*}

While varying $u$ acts on the propensity of individual nodes to participate in groups, the affinity matrix $w$ controls the density of interactions within and between communities. The generative model in \cref{eq: factorized prob} is well-defined for any non-negative symmetric affinity matrix $w$, allowing simulating various structures by properly tuning its entries. To illustrate the generation of hypergraphs with different affinity matrices, here we consider a range of matrices that start from diagonal (assortative) to gradually move to the uniform matrix of ones (disassortative), and rescale them to obtain an expected degree of five.
For simplicity we set the assignments $u$ to hard membership. The method is well suited to sample not only homogenous hypergraphs, but also higher-order networks with heterogenous distribution of the community size. Here we consider five communities with different sizes.
As shown in \cref{fig: affinity and community size}A, moving from an assortative to a disassortative configuration, the inter-community interactions strengthen substantially. Further, notice that the strengths of the interactions are influenced by the heterogeneity of the community size, as larger communities are expected to participate in more interactions.

It is also possible to tune individual entries of the affinity matrix $w$. In particular, in \cref{fig: affinity and community size}B we perform an experiment where we start from a diagonal matrix, and gradually increase only the $w_{12}$ (and $w_{21}$) entries, using three equally-sized communities. 
In this way, only the expected interactions between communities $1$ and $2$ are affected, while interactions among other communities are left unchanged. 

\subsection{Analzying community detection}
\label{sec: benchmarking community}
\begin{figure}[t]
\includegraphics[width=0.47\textwidth]{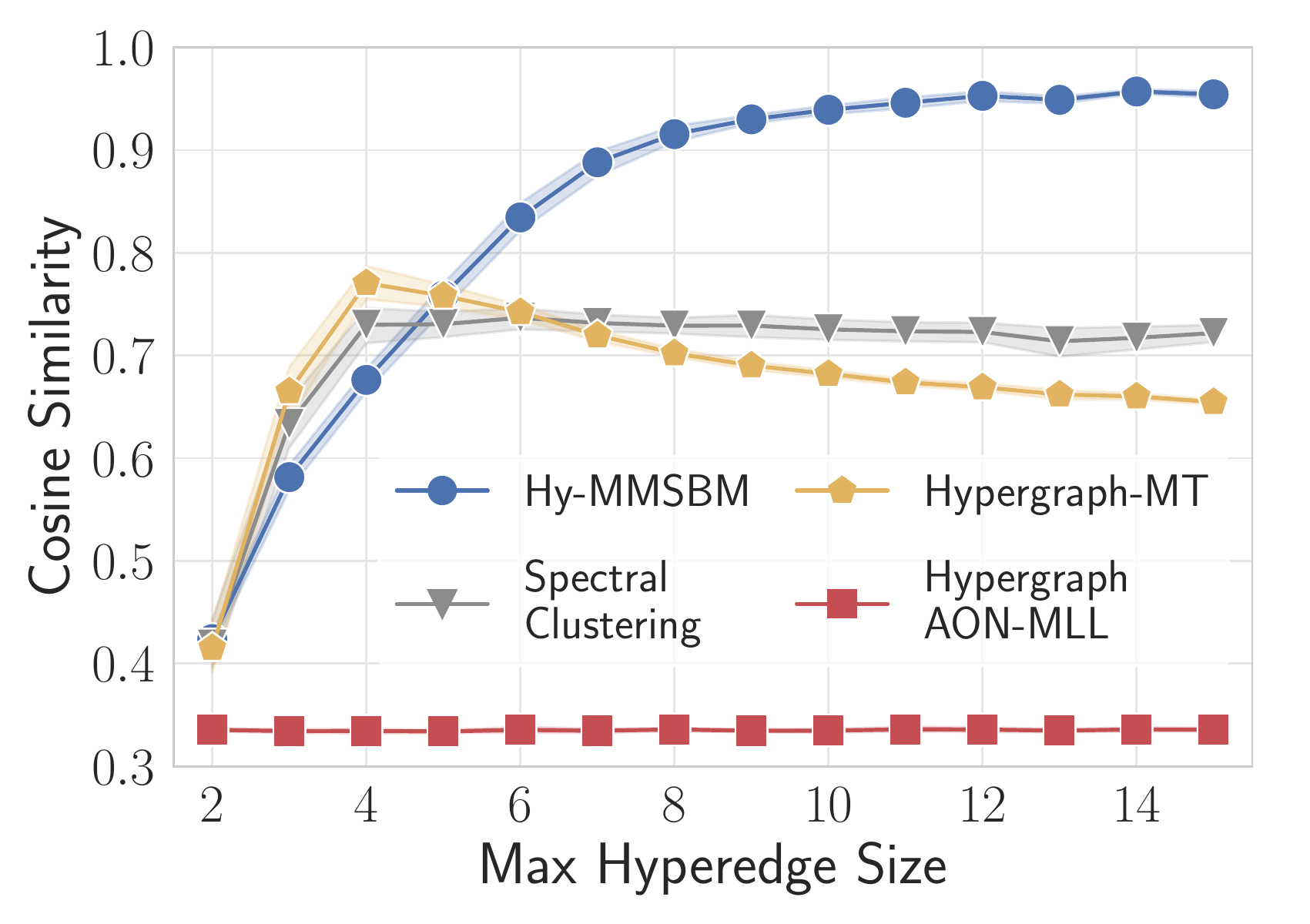}
\caption{\textbf{Evaluating higher-order community detection algorithms}.
    We sample hypergraphs to test the ability of different higher-order community detection algorithms to recover well-defined planted partitions. 
    We consider hypergraphs with $N=500$ nodes, $K=3$ equally sized assortative communities and hard assignments. We plot the cosine similarity between the inferred partitions and the ground truth as a function of the maximum hyperedge size. 
   Additional details on the data generation are given in  \cref{sec supp: generation of synthetic benchmark}.
}
\label{fig: inference on benchmark}
\end{figure}

One of the most useful applications of generating synthetic data with a desired underlying structure is the possibility to evaluate how competing algorithms perform on a given task that depends on the structure under control. In fact, when synthetic data with a known structure is available, it is possible to quantitatively compare the outcome of various algorithms and measure their ability to recover ground truth information. In network science, a classical and much investigated problem is assessing the ability of community detection algorithms to extract meaningful partitions of the network~\cite{lancichinetti2008benchmark}. For higher-order networks, the current lack of sampling methods for synthetic data with flexible community structure has led to a variety of custom-built examples, which renders comparison difficult and subject to individual choices~\cite{chodrow2021generative,chodrow2022nonbacktracking,contisciani2022principled}. 
 
In this section, we show how our synthetic data can be utilized to analyze the behavior of some of the current algorithms for higher-order community detection. To this end, we generate hypergraphs with assortative structure and hard community assignments, and perform inference with a variety of methods, namely \algoname{}~\cite{ruggeri2022inference}, Hypergraph-MT~\cite{contisciani2022principled}, spectral clustering~\cite{zhou2006learning} and hypergraph modularity~\cite{chodrow2021generative}.
In \cref{fig: inference on benchmark}, we show the cosine similarity of the inferred communities with the ground truth as a function of the maximum hyperedge size. As can be observed, \algoname{} attains the best performance when group interactions beyond a critical size are introduced, successfully recovering the ground truth assignments. \algoname{} is a flexible inference tool whose inference procedure is based on the same generative model described in \cref{eq: factorized prob}, and generally able to extract mixed-membership assignments for arbitrary (e.g. assortative or disassortative) community structure. Other algorithms attain varying scores, which might be explained by the different assumptions of the underlying models. For example, Hypergraph-MT is designed to extract overlapping communities, while spectral clustering can only be utilized for the detection of hard assignments. As such, the latter can be expected to perform well only in scenarios where interactions are dictated by hard communities, while the former can be employed when nodes may belong to more than one module. 

Procedures like the one presented in this section can be used to understand the limitations and strengths of different algorithms, allowing researchers to effectively test new proposals in different scenarios by varying the properties of the samples generated with our method, e.g. the degree of assortativity.

\subsection{Computational cost}
\begin{figure}[t]
\hspace{-1cm}
\includegraphics[width=0.42\textwidth]{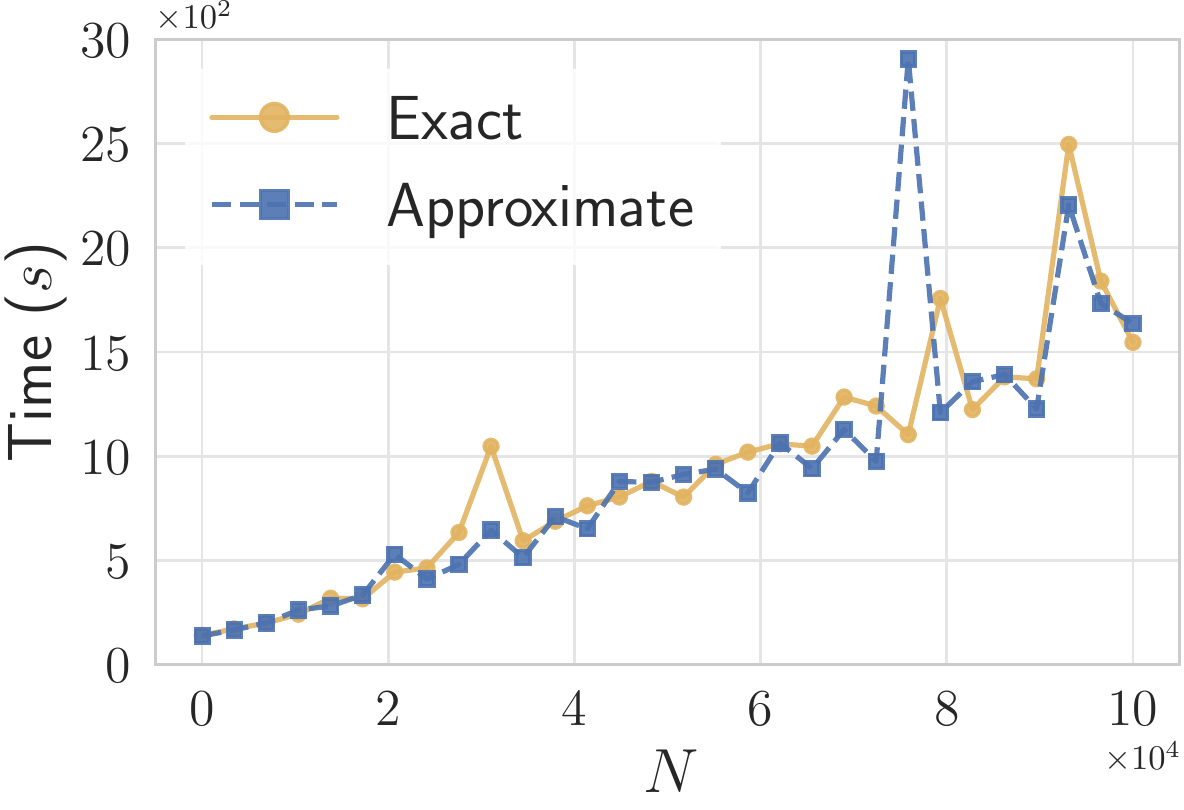}
\caption{
    \textbf{Computational complexity and scalability}. 
    We plot the computational cost of our sampling model for sparse hypergraphs as a function of the system size $N$. Our model is highly efficient, as it allows sampling of sparse hypergraphs of dimensions up to $N=10^5$ nodes in less than one hour. We show results for hypergraphs with fixed expected degree equal to 5, both for an exact (solid line) and an approximate approach (dashed line) based on central limit theorem sampling of dyadic interactions. Here, we utilize $K=5$ communities and unconstrained maximum hyperedge size $D=N$.
}
\label{fig: computational cost}
\end{figure}

Our sampling method is highly efficient and computationally scalable. We analyze the cost of our sampling strategy by discussing the cost of the individual sampling steps. 
The first step, consisting of sampling the degree and size sequences, can be cheaply performed in $O(N)$ time. In fact, to sample the $\degseq, \dimseq$ sequences we need to compute the mean and standard deviations defined in the Central Limit Theorem, and thus draw the sequences from the relative Gaussian distributions. These operations have linear cost, see \cref{sec supp: sampling step 1}.
In the second step we first combine the sampled $\degseq, \dimseq$ sequences into a first hyperedge configuration, and successively mix the hyperedges via MCMC. 
Generally, while the number of Markov chain steps needed for mixing is a function of $N$ and $|E|$ \cite{dutta2022sampling}, it is difficult to specify a pre-defined number.
In \cref{fig: computational cost}, we fix $n_b=100000$ burn-in steps and $n_i=20000$ intermediate steps between samples, which is a default value we utilized in most experiments. Nonetheless, the main cost we observe in this case is that prior to MCMC, i.e. the producing the first hyperedge configuration from the sequences. Empirically, such step dominates the computational cost.
Finally, the third step consists of sampling the non-zero weights according to $p(\w | \w^b; u,w)$. The cost of this operation is proportional to the number of hyperedges $|E|$; for sparse hypergraphs\textemdash and as often observed in real data\textemdash this is comparable to $N$. 

Empirically, we find the CLT approximations to be working well. 
Nevertheless, one could further improve on the quality of sampling by drawing the pairwise edges from their \textit{exact} Poisson distribution (\cref{eq: factorized prob,eq: lambda}), with cost $O(N^2)$, and resorting to approximations only for interactions of order three or greater. This is of particular help when sampling denser hypergraphs: since the MCMC does not necessarily guarantee non-repeated hyperedges, sampling directly the order-two interactions reduces the probability of repeated edges. For higher-order interactions, the probability of repetitions is negligible, in particular in sparse regimes \cite{chodrow2020configuration}.
Indeed, in all the experiments presented in this paper we sample the order-two interactions directly, and resort to the CLT approximations for hyperedges of order at least three.

In \cref{fig: computational cost} we investigate the efficiency of both exact and solely CLT-based sampling and observe the difference to be negligible. As discussed above, this is a consequence of the higher computational effort required in other sampling steps. Altogether, our model is highly efficient, as it allows sampling sparse hypergraphs of dimensions up to $10^5$ nodes in less than one hour.

\section{Real Data}
\label{sec:real data}

\subsection{Modeling real-world systems}
In this section we aim at sampling hypergraphs that mimic the community structure of a given dataset. To this end, we proceed as follows. First, we infer the affinity matrix $w$ and community assignments $u$ using the Hypergraph-MT algorithm \cite{contisciani2022principled} on the real data. Since this algorithm returns a (diagonal) matrix $w_d$ for every possible hyperedge size $d$, we take their element-wise geometrical mean to construct the matrix $w$ utilized in \cref{eq: lambda}.  
Notice that a similar approach could have been taken utilizing the Hy-MMSBM algorithm, which employs the same probabilistic model of our sampling method, as explained in \cref{sec: benchmarking community}. To highlight the flexibility of our methodology, which can be applied along with any community detection methodology, here we utilize Hypergraph-MT. In fact, our method accepts input parameter $w$ and $u$ regardless how these are obtained; in particular, these can be obtained by using different inference methods applied to the input data. Our method is capable of generating synthetic data conditioning on the desired input communities and affinity matrix. As such, it can be used in a complementary way together with community-based method focusing solely on inference.
Second, we condition the degree and size sequences by providing in input the observed hyperedge configuration, i.e. the hyperedges present in the real data. As explained in \cref{sec: additional user input}, this means skipping the first step of our sampling procedure and moving directly to perform MCMC starting from such configuration. The returned hypergraphs will have a structure similar to that of the data, but will be sampled according to the generative model in \cref{sec:generative model}.

\subsection{Comparing data and sample statistics}
\label{sec: data vs sample stats}
\begin{figure*}[t]
\centering
\includegraphics[width=\textwidth]{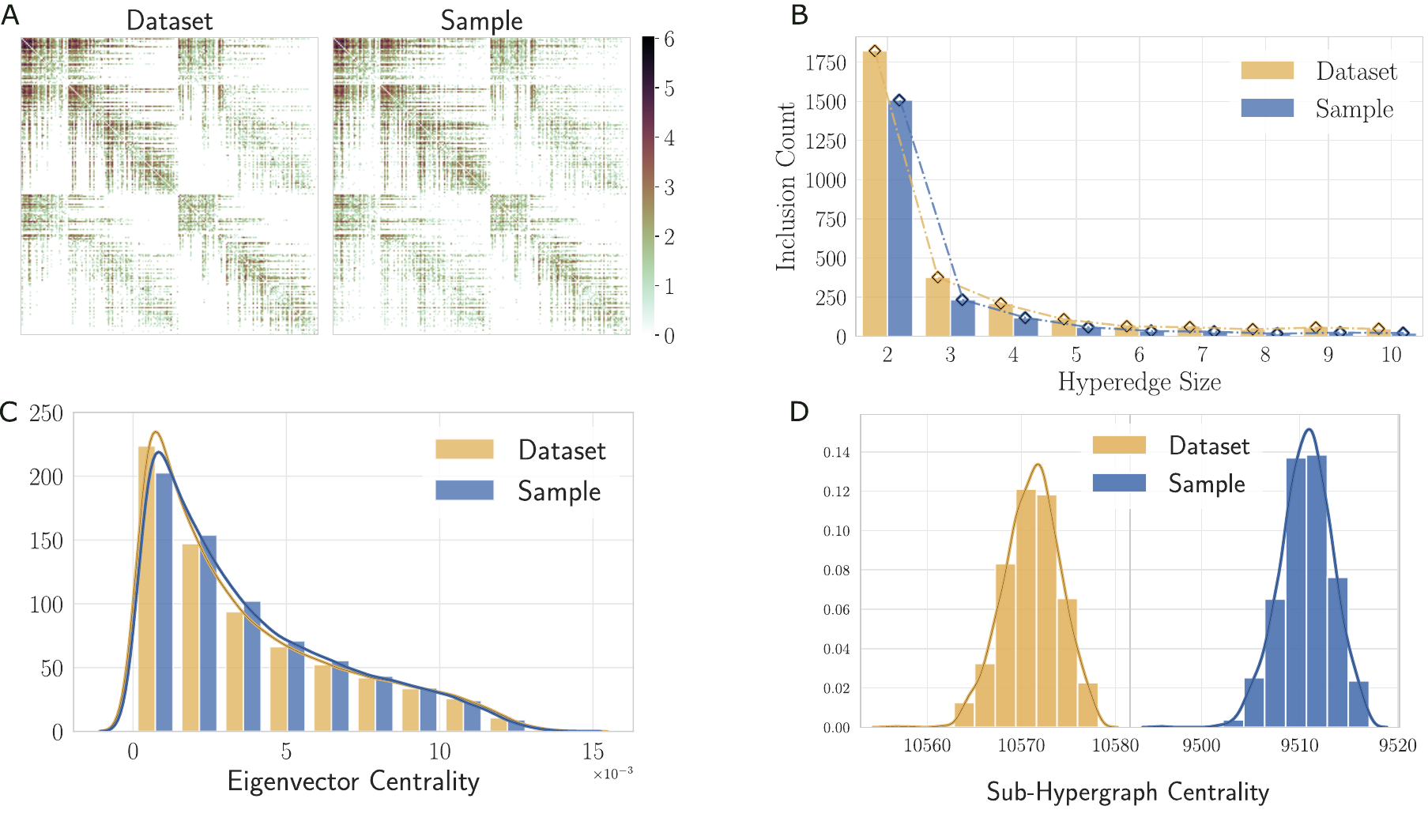}
\caption{
\textbf{Matching statistics of real-world data and samples: the case of the House Bills dataset} We plot (A) the adjacency matrices, (B) the hyperedge inclusions occurences, (C) the hyperedge eigenvector centrality distribution and (D) the sub-hypergraph centrality distribution for the House Bills dataset, where nodes represent congresspersons, and hyperedges describe subsets of them that co-sponsor a bill. For all such cases, we observe a good correspondence between the statistics measured on the real data and those obtained from a single sample of our generative model.
}
\label{fig: house bill results}
\end{figure*}

\begin{figure*}
\centering
\includegraphics[width=\textwidth]{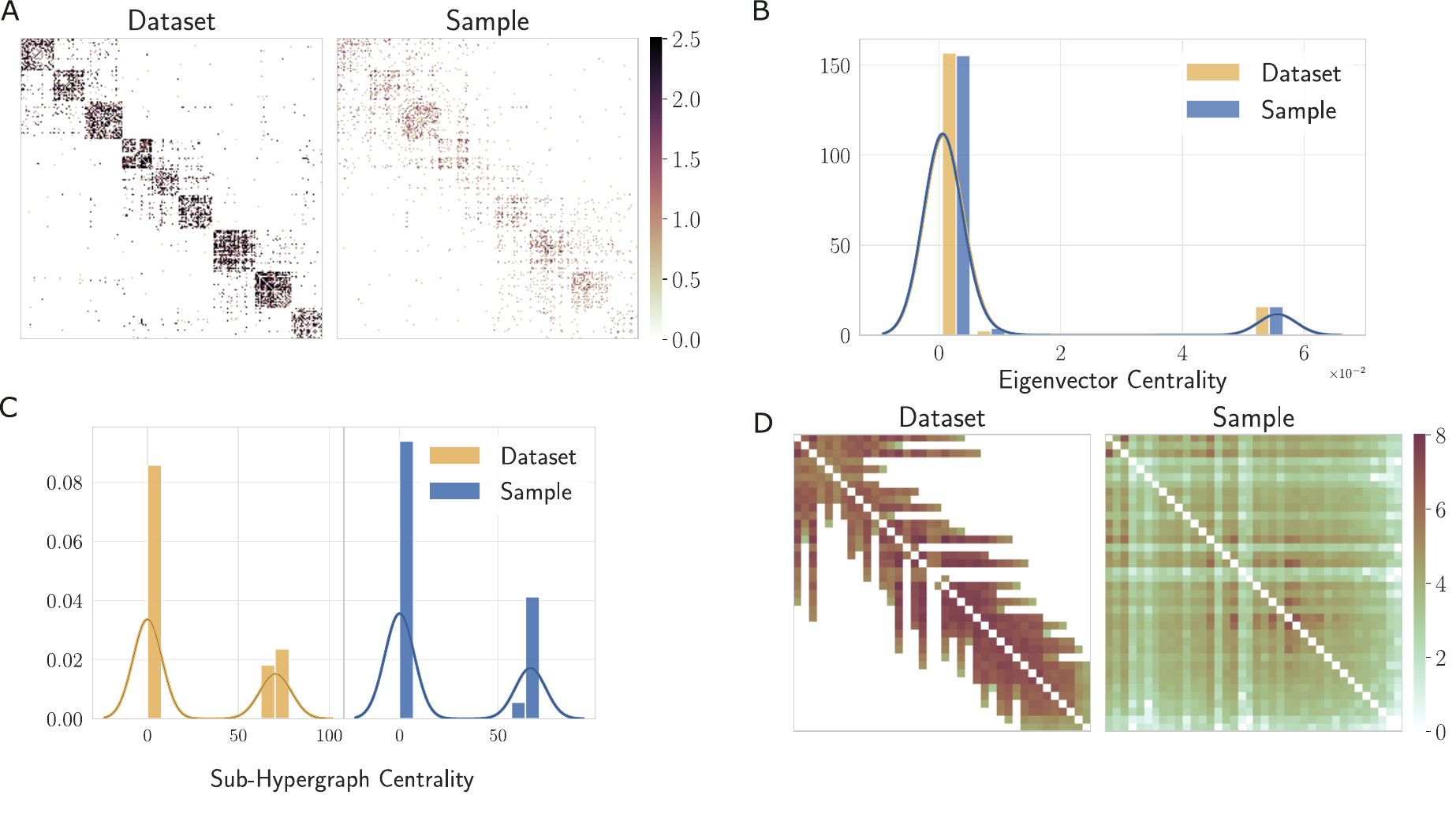}
\caption{
    \textbf{Hypergraph sample statistics, null hypothesis and generative assumptions}. 
    To illustrate the wide applicability of our model, we compare several statistics on real and sampled data. We plot (A) the adjacency matrix of face-to-face higher-order interactions among students in a High School dataset, (B) the eigenvector centrality distribution of co-purchasing behavior at Walmart, (C) the sub-hypergraph centrality distribution from committees data in the U.S. House. Similarly to the results presented in \cref{fig: house bill results}, our model correctly reproduces the desired statistics. In (D) we show the adjacency matrix associated with co-voting Justices of the U.S Supreme Court. Such data have a strong temporal structure which is not included in the generative assumption of the model, hence explaining the limited correspondence between real and synthetic statistics.}
\label{fig: other datasets}
\end{figure*}
We now apply the proposed methodology on a variety of real datasets. As a representative example, we consider a dataset of co-sponsoring of bills for the U.S. House of Representatives \cite{fowler2006connecting,fowler2006legislative}. Nodes correspond to congresspersons, and hyperedges connect subsets of them that co-sponsor a bill. The dataset contains $N=1494$ nodes, $|E|=54933$ hyperedges with maximum size $D=399$, and has been previously analysed via higher-order stochastic block models \cite{contisciani2022principled,chodrow2021generative}.

As a first sanity check, in Supp.Mat.~\cref{fig: sequences house bills} we verify that the degree and size sequences measured on the samples are identical to those of the data. This is guaranteed by the properties of the reshuffling operator described in \cref{sec: sampling}. 
We then proceed by comparing additional relevant statistics as measured on the real data and on the samples. Such statistics serve as a test for the goodness of fit, as they should match if the dataset is well-represented by the model.

We start by performing a visual comparison of the adjacency matrices \cite{estrada2006subgraph,battiston2020networks}, where the adjacency value $X_{ij}$ of any two nodes $i,j$ is defined as 
$X_{ij}:=\sum_{e\in E: i,j \in e}A_e \,$. As shown in \cref{fig: house bill results}A, our samples are well aligned with the real data.  

Another relevant structural property of a hypergraph is the inclusion relationships between hyperedges, i.e. which hyperedges are subsets of others~\cite{lotito2022higher}. This is of particular interest when comparing a hypergraph with its clique expansion, i.e. the graph obtained by projecting hyperedges onto pairwise interactions, or when comparing with other higher-order representations such as simplicial complexes~\cite{zhang2022higher,baccini2022weighted}. In \cref{fig: house bill results}B,  we count the number of hyperedges of size $n$ that are included in hyperedges of size $n+1$. Also in this case, results on our sample match well those measured on the input dataset.

Finally, we explore two centrality measures on hypergraphs. As a first example, in \cref{fig: house bill results}C we consider a generalization of eigenvector centrality \cite{bonacich1972factoring} for hyperedges. In particular, we consider the dual representation of the hypergraph, where nodes represent interactions in the original hypergraph and are connected if they have a non-empty intersection \cite{bretto2013hypergraph}. Moreover, in \cref{fig: house bill results}D we also compute sub-hypergraph centrality \cite{estrada2005complex,estrada2006subgraph}, which returns a measure of node importance in hypergraphs. Also in such cases, the quantities measured on our samples behave similarly to those based on the input data.

    We highlight that the resemblance between samples and real data is not simply due to the Markov Chain being stuck in a local optimum given by the initial configuration, i.e. the real dataset. To prove this, we further investigate the Markov Chain mixing while producing the samples based on the House-Bills dataset. We observe that $73\%$ of the shuffling steps are accepted by the Metropolis-Hastings algorithm, signalling good mixing. As an additional structural confirmation, we measure the Jaccard similarity between the real data and $10$ samples, defined as the number of hyperedges in the intersection divided by the number of hyperedges in the union. Also in this case, the resulting score of $0.69 \pm 0.11$ signals that the microscopic structure of the samples  detaches from that of the real data, while the macroscopic statistics in \cref{fig: house bill results} are preserved.
    Finally, we also observe that less structured methods fail to replicate such statistics. In \cref{sec supp: experiments configuration model} we obtain samples utilizing the configuration model from Chodrow~\cite{chodrow2020configuration}, which only takes into account the degree and size sequences. In this case, we observe a significant difference between the samples and the data, which could be explained by the lack of additional probabilistic structure in the sampling procedure.

To illustrate the wide applicability of our method, we extend this analysis to additional systems. In \cref{fig: other datasets}A we report the observed adjacency matrix of face-to-face interactions among High School students \cite{mastrandrea2015contact}, and the one obtained from a sample of our generative model. In \cref{fig: other datasets}B we show the distribution of the hyperedge eigenvector centrality computed on co-purchasing customer Walmart data \cite{amburg2020clustering}. Finally, in \cref{fig: other datasets}C we compare the sub-hypergraph centrality on the House Committees dataset \cite{chodrow2021generative,stewart2008congressional}, where hyperedges connect the members of the U.S. House participating in the same committees. In all such cases, we observe that our sampling method successfully models the desired statistics of the real data. 

Synthetic data generated to incorporate a particular structure are often utilized as tests for null hypotheses.
Indeed, discrepancies between sampled and real data may arise if some data features are not explicitly taken into account by the generative assumptions of the model \cite{hunter2008goodness}. Observing such differences can help unveil some relevant additional structure present in the data and originally neglected. As an example, we consider a dataset of co-voting patterns of the US Supreme Court Justices, where the nodes are Justices and hyperedges describe co-voting behaviors observed from 1946 to 2019 \cite{justicedata}. Since the number of Justices is fixed to 9 at any point in time, only interactions between Justices working in overlapping years can exist. Such an intrinsic time dependency, however, is not enforced by our model.
Hence, we do not expect samples of our model to match the input adjacency matrix well. We illustrate this in \cref{fig: other datasets}D, where the comparison of the sampled and observed adjacency matrices are distinctively different, with the real data showing a clear time-dependence. 
Our example illustrates the importance of correctly identifying the existence of particular structures in real-world dataset, showcasing how our sampling method could be used for testing null hypotheses and reproducing real-world statistics.

\section*{Discussion}
\label{sec:conclusions}
In this paper, we presented a framework for the generation of synthetic hypergraphs with flexible structure. 
Our model allows specifying different assortative and disassortative mesoscale configurations, tuning the size of the different communities and controlling the strengths of the interactions among them. Moreover, it allows regulating different node-level statistics, including hard or mixed community assignments and expected degrees. Through a variety of experiments, we showed how desired characteristics specified via input parameters are reflected in the generated data. Furthermore, we illustrated how practitioners can use our framework on real systems, both as a computationally efficient sampling tool for the replication of statistical measures, and as a structured null model for hypothesis testing. As an example, our model generates synthetic samples that successfully replicate centrality measures and inclusions relationship between hyperedges in higher-order data from different domains. Similarly, our model can help reveal important missing features in the generative assumptions made by different algorithms, showing clear discrepancies between samples and real data when, for instance, time-dependence is ignored. Finally, our framework allows testing the performance of different higher-order community detection methods.

There are various interesting and relevant avenues for future work. A first one is moving from the likelihood in \cref{eq: factorized prob}, which is based on a bilinear form, to one based on a multilinear form. While in principle this would allow for more flexible specifications, such as preventing the formation of certain hyperedge configurations, it is currently unclear how to obtain efficient expressions for the expected statistics and compute the moments required in the Central Limit Theorem. 
Moreover, additional information, such as time dependency and attributes on the nodes and hyperedges, could be explicitly incorporated in the probabilistic model. Such an extension could be based on insights from models for dyadic interactions, and result in substantial improvements when this information correlates with the hypergraph structure \cite{contisciani2020community,newman2016structure,zhang2017random,safdari2022reciprocity}.

Taken together, our methodology provides a principled, scalable and flexible framework to sample structured hypergraphs. 
To facilitate its usage we provide an open-source implementation at \repolink. The method is also implemented as part of the HGX library~\cite{lotito2023hypergraphx}.

\section*{Acknowledgements}
We thank Martina Contisciani for the extensive discussions and useful feedback. 
N.R. acknowledges support from the Max Planck ETH Center for Learning Systems.
F.B. acknowledges support from the Air Force Office of Scientific Research under award number FA8655-22-1-7025.

\newpage
\appendix
\section{The probabilistic model}
\label{sec supp: prob model}
We introduce some additional notation to that utilized in \cref{sec:generative model}. 
Recall that the hyperedges are independent realizations with distribution
\begin{equation*}
p(A_e; w, u) = \pois\left(A_e ; \frac{\lambda_e}{\normconst_{|e|}} \right) \quad \forall e \in \Omega \, ,
\end{equation*}
where we define
\begin{equation}\label{eq: lambda form supp}
\lambda_e := \sum_{i < j \in e} u_i^T w u_j \, .
\end{equation}
To avoid clutter, we overload the notation and define, for any hyperedge $e$,  $\normconst_e := \normconst_{|e|}$.\\
Furthermore, recall that $\Omega$ is the space of all possible hyperedges of sizes from $2$ to $D$. We also define, for any hyperedge size $d$, the space of hyperedges of fixed size $\Omega^d$.
The function $\delta$ is the indicator function, taking value $1$ if its argument is true, $0$ otherwise.

\subsection{Expected statistics}
\label{sec supp: expected stats}
Our model allows to obtain closed-form expressions for the expectations of various relevant statistics. In the following, we assume that $u, w$ are fixed, show how to derive some of these statistics and compute them in cheap linear time $O(N)$.
As explained in \cref{sec: additional user input}, having these statistics available is useful to aid the tuning of $u, w$ prior to sampling. We discuss the choice of the functional form of $\normconst$ in \cref{sec supp: kappa}.

\paragraph{Expected weighted degree of a node.}
We define the weighted degree $\degseq_i^w$ of a node $i$ as the weighted number of hyperedges it belongs to \cite{battiston2020networks}, that is:
\begin{equation*}
\degseq_i^w := \sum_{e \in E: i \in e} \w_e = \sum_{e \in \Omega: i \in e} \w_e  \, ,
\end{equation*}
due to the fact that $\w_e = 0$ for non-existing hyperedges $A_e \in \Omega \setminus E$.
Since $A_e$ is a random variable, the degree of node $i$ is also random and has expectation
\begin{align}
\mathbb{E}[d_i^w]
	&= \sum_{e \in \Omega: i \in e} \mathbb{E}[\w_e]= \sum_{e \in \Omega: i \in e} \frac{\lambda_e}{\normconst_e}	\nonumber \\
	&= \sum_{e \in \Omega: i \in e} \frac{1}{\normconst_e}
		\left(
			\sum_{j \in e: j \neq i} u_i^T w u_j
			+ \sum_{j <m \in e: j, m, \neq i} u_j^T w u_m
		\right) \nonumber \\
	&= \sum_{e \in \Omega: i \in e} \frac{1}{\normconst_e} \sum_{j \in e: j \neq i} u_i^T w u_j \nonumber \\
	&\hspace{4mm} + \sum_{e \in \Omega: i \in e} \frac{1}{\normconst_e} \sum_{j <m \in e: j, m, \neq i} u_j^T w u_m \nonumber \\
	&= \sum_{j \in V: j \neq i} 
		\left[
			\sum_{n=2}^D \frac{\binom{N-2}{n-2}}{\normconst_n}
		\right] u_i^T w u_j  \nonumber \\
	&\hspace{4mm}+	\sum_{j < m \in V: j, m \neq i} 
		\left[
			\sum_{n=3}^D \frac{\binom{N-3}{n-3}}{\normconst_n}
		\right] u_j^T w u_m \nonumber \\
	&= \left[
		\sum_{n=2}^D \frac{\binom{N-2}{n-2}}{\normconst_n}
	\right]	
	\left[
		 u_i^T w \bup{\sum_{j \in V: j \neq i} u_j}
	\right] \nonumber \\
	& \hspace{4mm} + 
	\left[
		\sum_{n=3}^D \frac{\binom{N-3}{n-3}}{\normconst_n}
	\right]	
	\left[
		\sum_{j < m \in V: j, m \neq i} u_j^T w u_m
	\right] \, .  \label{eq supp: expected degree of node}
\end{align}

The step from the fourth to fifth row is justified by counting the number of hyperedges of every size $n$ (normalized by the relative $\normconst_n$) where both nodes $i$ and $j$ are contained.  
Notice that the second summand $\sum_{j < m \in V: j, m \neq i} u_j^T w u_m$ has computational cost of $O(N^2)$.  We can reduce this to $O(N)$ by making the following general observation, which will also be used in other derivations. \\
For any fixed set of nodes $S$ and defining $s := \sum_{j \in S} u_j$:
\begin{align}
\sum_{j < m \in S} u_j^T w u_m
	&= \frac{1}{2} \sum_{j, m \in S,  j \neq m} u_j^T w u_m \nonumber \\
	&= \frac{1}{2} \left(
		\sum_{j, m \in S} u_j^T w u_m - \sum_{j \in S} u_j^T w u_j
	\right) \nonumber \\
	&= \frac{1}{2} \left(
		\sum_{j \in S} \sum_{m \in S} u_j^T w u_m - \sum_{j \in S} u_j^T w u_j
	\right) \nonumber \\
	&= \frac{1}{2} \left(
		s^T w s - \sum_{j \in S} u_j^T w u_j
	\right) \, . \label{eq supp: linear trick}
\end{align}
Both these terms can be calculated in $O(|S|)$. \\
In the case of the expected degree of node $i$, the second summand of \cref{eq supp: expected degree of node} can be computed with $S = V \setminus \{i\}$, while the first summand can be directly computed in linear time.

\paragraph{Expected weighted degree.}
This is the average weighted degree of all the nodes in the network,  and is given by
\begin{align}
\langle d^w \rangle
	&= \frac{1}{N} \sum_{i \in V} \mathbb{E}[d_i^w] 
            = \frac{1}{N} \sum_{i \in V} \sum_{e \in \Omega: i \in e} \frac{\lambda_e}{\normconst_e} \nonumber \\
	&= \frac{1}{N} \sum_{e \in \Omega} \sum_{i \in e} \frac{\lambda_e}{\normconst_e} 
            = \frac{1}{N} \sum_{e \in \Omega}\, |e|\, \frac{\lambda_e}{\normconst_e}  \nonumber \\
	&= \frac{1}{N} \sum_{e \in \Omega} \frac{|e|}{\normconst_e} \sum_{i < j \in e} u_i^T w u_j \nonumber \\
	&= \frac{1}{N} 
		\left(
			\sum_{n=2}^D \binom{N-2}{n-2} \frac{n}{\normconst_n}
		\right)
		\sum_{i < j \in V} u_i^T w u_j \, .  \label{eq: expected deg formula}
\end{align}
This quantity can be reduced to $O(N)$ cost by utilizing the trick in \cref{eq supp: linear trick}.

\paragraph{Accounting only for specified interactions} 
The statistics described above can also be computed by taking into account only interactions of a fixed size.  For example, a user might be interested in computing the expected degree of a node by considering only hyperedges of sizes up to a certain value, or only for a fixed hyperedge size. These can be readily computed by repeating the derivations above.
For example, computing the expected degree in Eq.~\eqref{eq: expected deg formula} only for hyperedges of sizes $2, 3$ or $4$, we obtain
\begin{equation*}
\frac{1}{N} 
		\left(
			\sum_{n=2}^4 \binom{N-2}{n-2} \frac{n}{\normconst_n}
		\right)
		\sum_{i < j \in V} u_i^T w u_j \ .
\end{equation*}
Notice that only the multiplicative constant $\sum_{n=2}^4 \binom{N-2}{n-2} \frac{n}{\normconst_n}$ has changed. 

\subsection{Choosing the normalization $\normconst_n$}
\label{sec supp: kappa}
The normalization constant $\normconst_n$ rescales the probabilities of the hyperedges of size $n$.  
This rescaling is needed to contrast the effects of the high-dimensional configuration space $\Omega$. Removing the constant (i.e. setting $\normconst_n \equiv 1$ for all $n$) yields exploding statistics due to the combinatorial factors appearing for larger hyperedges, see e.g. \cref{eq: expected deg formula}.

While it is theoretically possible to sample from a model with such $\normconst_n$ values, the expected degree and size sequences would not match those observed in real data. In all our experiments we choose instead the following form for $\normconst_n$, in a way that yields reasonable expected statistics:  
\begin{equation}\label{eq: default kappa}
\normconst_n := \frac{n (n-1)}{2}\binom{N-2}{n - 2} \, .
\end{equation}
This expression satisfies two important properties. First, $\kappa_2 = 1$, so that the probabilistic model restricted to binary interactions is equivalent to the standard Poisson stochastic block model, second, the expected degree in \cref{eq: expected deg formula} reduces to 
\begin{equation*}
\langle d^w \rangle = \frac{1}{N} 
		\left(
			\sum_{n=1}^{D-1} \frac{1}{n}
		\right)
		\sum_{i < j \in V} u_i^T w u_j \, .
\end{equation*}
This avoids combinatorial explosions in the expected degree, allowing to tune the model based only on $u, w$. \\
The form~\eqref{eq: default kappa} has also a valid interpretation. The binomial $\binom{N-2}{n - 2}$ normalizes for the number of possible hyperedges of size $n$ that any two fixed nodes belong to (since one needs to choose the remaining $n-2$ nodes in the hyperedge among the possible $N-2$). The value $\frac{n (n-1)}{2}$ is the number of possible binary interactions among $n$ nodes, and is used to take the average of the summands appearing in the expression~\eqref{eq: lambda form supp} for the sufficient statistics: $\lambda_e = \sum_{i < j \in e} u_i^T w u_j$. We also note that similar combinatorial expressions arise naturally in the literature, due to the exploding configuration space \cite{chodrow2020configuration,pal2021community}.

\noindent
While practitioners can make other possible choices with similar properties, e.g.  $\normconst_n = \binom{N-2}{n-2}$ or $\normconst_n = \frac{2}{n}\binom{N-2}{n-2}$, we remark that the methodology and the theory proposed in this paper hold for any choice of $\normconst_n > 0$.

\section{Technical details about sampling}
\label{sec supp: sampling}
We include here all the technical details to approximately sample from the probabilistic model. 
We start by giving some definitions, then we proceed to outline the three sampling steps introduced in \cref{sec: sampling}.
In the following, we consider any fixed choice of parameters $u, w$. Notice also that all the probabilities $p(\cdot)$ utilized in this section depend on $u,w$. 
To avoid clutter, we implicitly assume this dependency in the following derivations.

\begin{definition*}
The \emph{unweighted} (or \emph{binary}) \emph{hypergraph} is the hypergraph with $\w^b \in \{0, 1\}^{|\Omega|}$ derived from the original weights $\w \in \mathbb{N}^{|\Omega|}$.  Formally:
\begin{equation*}
\w^b_e := \delta(\w_e > 0) \,,\quad \forall e \in \Omega \, .
\end{equation*}
The \emph{binary degree sequence} $\degseq$ is the degree sequence in the binary hypergraph.  In other words,  it is the degree sequence in the original hypergraph if we consider the hyperedges as unweighted.
\end{definition*}

The sampling procedure is divided in three consecutive steps:
\begin{itemize}
\item First,  approximately sample the binary degree and size sequences $(\degseq,  \dimseq)$.
\item Second,  sample the binary hypergraph $\w^b$ from $p(\w^b | \degseq,  \dimseq)$.  Notice that, since we sample a binary hypergraph,  hyperedges can only exist or not, i.e. $\w^b_e \in \{0, 1\} \, \forall e \in \Omega$.
\item Third,  sample the weights of the final graph given the binary one: $p(\w | \w^b)$.
\end{itemize}

\noindent
\textit{Why is this correct?} We aim at sampling from $p(\w)$. The procedure above corresponds instead to sampling from $p(\w,  \w^b,  \degseq,  \dimseq)$,  so why is this correct? Notice that
\begin{align*}
\w &\text{ uniquely defines } \w^b \\
\w^b &\text{ uniquely defines } (\degseq, \dimseq) \, .
\end{align*}
Thus, if all the sampled quantities $\w, \w^b, \degseq, \dimseq$ are compatible, we can observe that both the following equalities are true:
\begin{align*}
p(\w,  \w^b,  \degseq, \dimseq)
	&= p(\w^b, \degseq,  \dimseq| \w) \, p(\w) \\
	&= p(\w) \\
p(\w, \w^b,  \degseq, \dimseq) 
	&= p(\w | \w^b,  \degseq, \dimseq) \, p(\w^b | \degseq, \dimseq) \, p(\degseq, \dimseq) \\
	&= p(\w | \w^b) \, p(\w^b | \degseq, \dimseq) \, p(\degseq,\dimseq) \, ,
\end{align*}
hence
\begin{equation*}
p(\w) = p(\w | \w^b) \, p(\w^b | \degseq, \dimseq)\, p(\degseq, \dimseq) \, .
\end{equation*}
Having verified the consistency of our sampling routine, we now proceed describing the three steps in more details.

\subsection{Sampling the binary degree and \\size sequences}
\label{sec supp: sampling step 1}
The sequences $\degseq, \dimseq$ cannot be sampled exactly and efficiently. We propose instead an approximation based on a version of the Central Limit Theorem. \\

\noindent
Consider the binary degree sequence $\degseq$. For a node $i$,  we need to sample the number $\degseq_i$ of existing hyperedges that $i$ belongs to:
\begin{align}
\degseq_i
	&:= \sum_{e \in \Omega :  i \in e} \delta(\w_e > 0) \, .
\end{align}  
Notice that the summands $\delta(\w_e > 0)$ are independent Bernoulli random variables with probability $p_e = 1- \text{Pois}({A_e=0;\lambda_e/\kappa_e}) = 1- \exp\left(- \frac{\lambda_e}{\normconst_e} \right)$,  therefore easy to sample one by one.  Due to the exponential size of $\Omega$, however,  sampling all of them is practically impossible.  \\

\noindent
Similarly, consider the size sequence $\dimseq$. For every hyperedge size $\ell \in \{2, \ldots, D\}$ (potentially up to $D=N$),  we need to sample the number $\dimseq_\ell$ of hyperedges of such size, defined as:
\begin{equation*}
\dimseq_{\ell} = \sum_{e \in \Omega^{\ell}} \delta(\w_e > 0) \,,
\end{equation*}
where $\Omega^{\ell}=\ccup{e \in \Omega\;\vert \: |e| = \ell}$.
\noindent
The following theorem helps in approximately sampling these quantities.
\begin{theorem}
\label{th:binary sequence approx}
Consider $\degseq_i,\dimseq_{\ell}$ as defined above.  Furthermore, assume that $u$ is bounded,  i.e. $\exists L > 0: u <L$, where the inequality is intended element-wise.  Then:
\begin{enumerate}[label={\alph*}.]
\item  For any hyperedge size $\ell \ge 3$, both $\degseq_i$ and $\dimseq_{\ell}$ satisfy the assumptions of the Lyapunov Central Limit Theorem (see \cref{supp sec: lyapunov clt} for the statement). Thus, they can be approximately sampled from a Gaussian. 
\item Furthermore,  if we assume that the assignments $u$ are lower bounded away from zero, i.e. $\exists \epsilon > 0 \text{ s.t. } u > \epsilon$, then the statement above also holds for $\ell=2$.
\item To a first-degree approximation, we can compute the mean and variance needed for the asymptotic Gaussian distributions as:
\begin{align}
\mathbb{E}[\dimseq_{\ell}] &\approx \text{Var}(\dimseq_{\ell}) \approx \sum_{e \in \Omega^{\ell}} \frac{\lambda_e}{\kappa_e} \label{eq: dimseq th expr} \\
\mathbb{E}[\degseq_i] &\approx \text{Var}(\degseq_i) \approx \sum_{e \in \Omega: i \in e} \frac{\lambda_e}{\kappa_e} \, ,  \label{eq: degseq th expr}
\end{align}
i.e. can be approximated by the weighted number of hyperedges and weighted degree.
\end{enumerate}
\end{theorem}
The proof can be found in \cref{sec supp: proof of theorem on binary seq approx}. Practically, the sampling proceeds as follows. First, we compute the theoretical values in \cref{eq: dimseq th expr,eq: degseq th expr}, this can be done in linear time similarly to the statistics in \cref{sec supp: expected stats}. Then, we separately sample $\degseq, \dimseq$ from Gaussian distributions with the given means and variances. Since both sequences $\degseq, \dimseq$ need to take integer values, we then round the samples element-wise to the closest integer. Finally, we combine the sampled sequences into a first list of hyperedges, representing a binary (i.e. unweighted) hypergraph. We describe the recombining algorithm in \cref{sec supp: merging sequences}.

\subsection{Sampling the binary hypergraph}
\label{sec supp: sampling step 2}
Once we condition on the binary degree and size sequences, we sample which hyperedges will be present in the final hypergraph, i.e. all $e\in \Omega$ such that $A_e^b=1$. \\
Formally, the conditional sampling of hyperedges can be performed via the MCMC procedure introduced in Chodrow et al.~\cite{chodrow2020configuration}. For this, we use a hyperedge reshuffling operator to define the MCMC steps yielding a valid Markov chain that preserves the initial $\degseq, \dimseq$. The Markov chain starts from a proposal list of hyperedges that is then modified at every step. Notice that the main difference with the MCMC procedure proposed in the original paper is that the acceptance-rejection probabilities are based on our generative models, and hence they depend on the $u,w$ parameters. Hence, we present next how to compute such probabilities in detail.

During the MCMC, we start from two hyperedges $e_1,e_2$ and shuffle them to form two new hyperedges $e_1',  e_2'$. Importantly, one of the properties of the shuffle operator is that $|e_1| = |e_1'|$ and $|e_2| = |e_2'|$. The Metropolis-Hastings transition probability to be computed is then:
\begin{align*}
	& \frac{
        p(\text{new  configuration})
}{ p(\text{current configuration})} \\
	&= \frac{
       p(A_{e_1} = 0)\, p(A_{e_2} = 0)\, p(A_{e_1'} = 1)\, p(A_{e_2'} = 1)	
	}{p(A_{e_1} = 1) \,p(A_{e_2} = 1) \,p(A_{e_1'} = 0)\, p(A_{e_2'} = 0)} \\
	&= 
	\frac{
	    \left(\exp\left(\frac{\lambda_{e_1'}}{\normconst_{e_1'}}\right) - 1 \right)
	    \left(\exp\left(\frac{\lambda_{e_2'}}{\normconst_{e_2'}}\right) - 1 \right)
	}{
        \left(\exp\left(\frac{\lambda_{e_1}}{\normconst_{e_1}}\right) - 1\right)
        \left(\exp\left(\frac{\lambda_{e_2}}{\normconst_{e_2}}\right) - 1\right)
	} \, .
\end{align*}
For some hyperedges it could happen that $\normconst_{e} \gg \lambda_{e}$, which may lead to numerical instabilities, as over or underflows in the exponentials (the same holds for the other hyperedges). To mitigate this risk we compute all the $\lambda$ and the $\normconst$ in log-space. Using the fact that $e^x-1\approx x$ when $x \ll 1$, if we measure that for a hyperedge $e$ we have $\log \kappa_e - \log \lambda_e  > \tau$ for a certain threshold $\tau$, then we apply the approximation
\begin{equation*}
\exp\left(\frac{\lambda_{e}}{\kappa_{e}}\right) - 1 \approx \frac{\lambda_{e}}{\kappa_{e}} \quad.
\end{equation*}
In practice, when one among $e_1$ and $e_1'$ (and similarly for $e_2, e_2'$) satisfies the above condition, then we approximate the following ratio as:
\begin{equation*}
\frac{
    \exp\left(\frac{\lambda_{e_1'}}{\normconst_{e_1'}}\right) - 1
}{
    \exp\left(\frac{\lambda_{e_1}}{\normconst_{e_1}}\right) - 1 
}
\approx
\frac{
    \left(\frac{\lambda_{e_1'}}{\normconst_{e_1'}}\right) 
}{
    \left(\frac{\lambda_{e_1}}{\normconst_{e_1}}\right) 
}
= 
\frac{\lambda_{e_1'}}{\lambda_{e_1}} \, ,
\end{equation*}
where the last step is due to the fact that  the hyperedges have the same size, i.e. $|e_1| = |e_1'|$. This avoids computing the $\normconst$ values.

\subsection{Sampling the weights of the hypergraph}
\label{sec supp: sampling step 3}
Finally,  we need to sample the Poisson weights from $p(\w | \w^b)$. Provided that the $\w^b$ weights yield a sparse realization, we only need to sample the weights of the few hyperedges present, signalled by $\w^b =1$.  
Hence, we need to sample from the distribution 
\begin{equation*}
\mathbb{P}(\w_e = n | \w^b_e=1) = \mathbb{P}(\w_e = n | \w_e > 0) \, ,
\end{equation*}
which is a zero-truncated Poisson distribution.  Sampling from such a distribution efficiently is not immediate, we propose a solution based on \textit{inverse transform sampling} \cite{ross2022simulation} next.

Consider a Poisson random variable $X \sim \pois(\lambda)$.
We aim at sampling from the distribution of the random variable $Y$ defined as 
\begin{equation*}
Y := X | X > 0 \, .
\end{equation*}
To this end, we compute the cumulative distribution function (cdf) of $Y$. For any $v \in \mathbb{N} \setminus \{0\}$
\begin{align*}
F_Y (v)
	&= \mathbb{P}(Y \le v) \\
	&= \sum_{m = 1}^v \mathbb{P}(Y = m) \\
	&= \frac{1}{1-e^{-\lambda}} \sum_{m = 1}^v \mathbb{P}(X = m) \\
	&= \frac{F_X(v) - \mathbb{P}(X=0)}{1-e^{-\lambda}}  \\
	&= \frac{F_X(v) - e^{-\lambda}}{1-e^{-\lambda}}  \, .
\end{align*}
Its inverse,  called inverse-cdf or percent-point-function, is given for any $p \in [0, 1]$ by 
\begin{align}
Q_Y(p)
	&= \min \left\{ v \in \mathbb{N} \, | \,  p \le F_Y(v) \right\} \nonumber \\
	&= \min \left\{ v \in \mathbb{N} \, | \,  p \le \frac{F_X(v) - e^{-\lambda}}{1-e^{-\lambda}} \right\} \nonumber \\
	&= \min \left\{ v \in \mathbb{N} \, | \,  e^{-\lambda} + p (1-e^{-\lambda}) \le F_X(v) \right\} \nonumber \\
	&= Q_X(e^{-\lambda} + p (1-e^{-\lambda})) \, . \label{eq: truncated poisson inverse cdf}
\end{align}
Thus, one can sample from $Y$ by drawing a uniform random variable $p \sim \text{Unif}(0, 1)$ and computing $Q_Y(p)$ via Eq.~\eqref{eq: truncated poisson inverse cdf}.
Crucially, this is cheap to compute (and simple to implement) because it corresponds to the inverse-cdf of a Poisson distribution.

\section{Proof of \cref{th:binary sequence approx}}
\label{sec supp: proof of theorem on binary seq approx}
\begin{proof}
The derivations for the binary degree and size sequences are very similar, for simplicity we only present a proof for the size sequence $\dimseq_{\ell}$, for any fixed possible value of $\ell$.  In the following, call $I_e := \delta(\w_e>0)$ and $p_e$ its Bernoulli probability $p_e = 1 - \exp\left(-\frac{\lambda_e}{\normconst_{\ell}}\right)$. 

\paragraph{Lyapunov CLT for $\ell \ge 3$} A possible way to show that the Lyapunov CLT applies, is to show that 
\begin{equation}\label{eq: lyapunov condition}
\lim_{N \rightarrow +\infty} \frac{
    \sum_{e \in \Omega^{\ell}} \mathbb{E}\rup{\left| I_e - \mathbb{E}[I_e] \right|^3}
}
{
    \left( \sqrt{\sum_{e \in \Omega^{\ell}} \text{Var}(I_e)  } \right)^3
} = 0 \, .
\end{equation}
First,  observe that:
\begin{align*}
\lim_{N \rightarrow +\infty} \frac{
    \sum_{e \in \Omega^{\ell}} \mathbb{E}\rup{ \left| I_e - \mathbb{E}[I_e] \right|^3}
}
{
    \left( \sqrt{\sum_{e \in \Omega^{\ell}} \text{Var}(I_e)  } \right)^3
}  \\
	&\hspace{-4cm}=
	    \lim_{N \rightarrow +\infty} \frac{
        \sum_{e \in \Omega^{\ell}} p_e (1 - p_e)\bup{(1-p_e)^2 + p_e^2}
    }
    {
        \bup{\sum_{e \in \Omega^{\ell}} p_e (1-p_e)}^{3/2}
    } \\
    &\hspace{-4cm}\le
	    2 \lim_{N \rightarrow +\infty} \frac{
        \sum_{e \in \Omega^{\ell}} p_e (1 - p_e)
    }
    {
        \bup{\sum_{e \in \Omega^{\ell}} p_e (1-p_e)}^{3/2}
    } \\
    &\hspace{-4cm}=
    2 \lim_{N \rightarrow +\infty} \frac{1}
    {
        \bup{\sum_{e \in \Omega^{\ell}} p_e (1-p_e)}^{1/2}
    } \, .
\end{align*}
Hence, proving that $\sum_{e \in \Omega^{\ell}} p_e (1 - p_e) \rightarrow +\infty$,  means that the Lyapunov condition in~\cref{eq: lyapunov condition} is satisfied.  Due to the assumption that the $u$ are bounded, there exists some $0 < L' < 1$ such that $p_e < L'$ for all $e$ (recall that $p_e$ depends on $u$ through $\lambda_e$ in~\cref{eq: lambda form supp}).  In the following derivation,  let $i, j \in V$ be two arbitrary nodes.  Then:
\begin{align*}
\sum_{e \in \Omega^{\ell}} p_e (1 - p_e) &\ge (1-L') \sum_{e \in \Omega^{\ell}} p_e \\
	&\hspace{-17mm}\ge (1-L') \sum_{e \in \Omega^{\ell}: i, j \in e} p_e \\
	&\hspace{-17mm}= (1-L') \sum_{e \in \Omega^{\ell}: i, j \in e} \rup{1- \exp\left( - \frac{\sum_{l < m \in e}  u_l^T w u_m }{\normconst_{\ell}}      \right) }\\
	&\hspace{-17mm}\ge (1-L') \sum_{e \in \Omega^{\ell}: i, j \in e} \rup{1- \exp\left( - \frac{u_i^T w u_j }{\normconst_{\ell}}\right)} \\
	&\hspace{-17mm}= (1-L') \rup{1- \exp\bup{ - \frac{u_i^T w u_j }{\normconst_{\ell}}}} \sum_{e \in \Omega^{\ell}: i, j \in e} \hspace{-2mm} 1 \, .
\end{align*}
Finally, notice that $\sum_{e \in \Omega^{\ell}: i, j \in e} 1 \rightarrow +\infty$ with $N \rightarrow + \infty$,  due to the fact that the number of hyperedges containing $i, j$ tends to infinity (if the hyperedge size $\ell$ is at least $3$).

\paragraph{Lyapunov CLT for $\ell=2$} Here we also assume that $u < \epsilon$. Then there exists some $L'' >0$ s.t.,  for any two nodes $m$ and $l$,  their interaction is bounded away from zero, i.e. $u_l^T w u_m > L''$.  Similarly to above,  let $i$ be an arbitrary node:
\begin{align*}
\sum_{e \in \Omega^{\ell}} p_e (1 - p_e)
&\ge (1-L') \sum_{e \in \Omega^{\ell}: i\in e} p_e \\
	&\hspace{-1.1cm}= (1-L') \sum_{e \in \Omega^{\ell}: i\in e} \rup{1- \exp\left( - \frac{\sum_{l < m \in e}  u_l^T w u_m }{\normconst_{\ell}}      \right)} \\
	&\hspace{-1.1cm}\ge (1-L') \sum_{e \in \Omega^{\ell}: i\in e}\rup{ 1- \exp\left( - \frac{L''}{\normconst_{\ell}}      \right) }\\	
	&\hspace{-1.1cm}= (1-L') \rup{1- \exp\left( - \frac{L''}{\normconst_{\ell}} \right) } \sum_{e \in \Omega^{\ell}: i\in e} 1  \, .
\end{align*}
Similar to the case for $\ell\ge3$, the quantity $\sum_{e \in \Omega^{\ell}: i\in e} 1$ tends to infinity for diverging values of $N$.

\paragraph{Approximation of the statistics}
The statistics $\mathbb{E}[\dimseq_{\ell}]$ and $\text{Var}[\dimseq_{\ell}]$ are hard to compute efficiently in closed form. However, we can approximate them using the fact that $1 -e^{-x} \approx x$ and $(1 -e^{-x})\,e^{-x} \approx x$, when $x \ll 1$. 
Then
\begin{align*}
\mathbb{E}[\dimseq_{\ell}] 
	&= \sum_{e \in \Omega^{\ell}} 1 - e^{- \frac{\lambda_e}{\normconst_{\ell}}} \approx \sum_{e \in \Omega^{\ell}} \frac{\lambda_e}{\normconst_{\ell}} \\
\text{Var}(\dimseq_{\ell})
	&= \sum_{e \in \Omega^{\ell}} \left(1 - e^{- \frac{\lambda_e}{\normconst_{\ell}}}\right) e^{- \frac{\lambda_e}{\normconst_{\ell}}} \approx \sum_{e \in \Omega^{\ell}} \frac{\lambda_e}{\normconst_{\ell}} \, .
\end{align*}
We expect these approximations to be valid when $\frac{\lambda_e}{\normconst_e} \ll 1$. This is a sparse regime attained when only few among all the possible hyperedges are present and is a realistic assumption in many practical applications.
\end{proof}

\section{Lyapunov Central Limit Theorem}
\label{supp sec: lyapunov clt}
We state here the Lyapunov Central Limit Theorem \cite{billingsley2008probability} that we utilize in our main proof in \cref{sec supp: proof of theorem on binary seq approx}.
\begin{theorem}
Consider a sequence of independent random variables $\{X_i\}_{i \in \mathbb{N}}$,  and define $\mu_i := \mathbb{E}[X_i]$,  and $\sigma^2_i := \text{Var}(X_i)$.  Also, define
\begin{equation*}
s_n^2 := \sum_{i=0}^n \sigma^2_i \, .
\end{equation*}
If there exists some $\epsilon > 0$ such that
\begin{equation*}
\lim_{n \rightarrow +\infty} \frac{1}{s_n^{2+\epsilon}} \sum_{i=0}^n \mathbb{E}\left[ \left| X_i - \mu_i	\right|^{2+\epsilon} \right] 	= 0 \, ,
\end{equation*}
then the quantity
\begin{equation*}
\frac{1}{s_n} \sum_{i=0}^n (X_i - \mu_i) 
\end{equation*}
converges in distribution to a standard Gaussian random variable.
\end{theorem}

\section{A matching algorithm for the $\degseq, \dimseq$ sequences}
\label{sec supp: merging sequences}
After separately sampling the binary degree sequence $\degseq$ and the size sequence $\dimseq$,  as described in \cref{sec supp: sampling step 1}, these need to be combined to obtain a valid set of hyperedges.  However,  given two arbitrary sequences $\degseq, \dimseq$,  there does not necessarily exist a hypergraph that satisfies both.  For example,  the average (unweighted) degree can be calculated both from $\degseq$ and $\dimseq$; the two values obtained need to match.  For this reason, we need a ``matching algorithm'' to extract a set of hyperedges that are consistent with both $\degseq, \dimseq$. Notice that this task in not straightforward. Indeed, also Chodrow~\cite{chodrow2020configuration}\textemdash who first introduced the Markov chain shuffling procedure\textemdash had to start the MCMC from a valid hyperedge list. The currently  available implementation \footnote{\href{https://github.com/PhilChodrow/hypergraph}{https://github.com/PhilChodrow/hypergraph}} can only replicate the sequences of existing hypergraphs, but it is not clear how to proceed from scratch with no initial data, thus reducing the applicability of the method. Our work generalizes the sampling to arbitrary starting sequences. This is a contribution in-and-of-itself, as in principle this could be applied to other sampling procedures with different underlying generative processes than the one we described in \cref{sec:generative model}, as long as they provide some initial $\degseq, \dimseq$ sequences.

Before presenting our method, we make some remarks. First, sampling separately $\degseq, \dimseq$ assumes the approximation $p(\degseq, \dimseq) \approx p(\degseq) \,p(\dimseq)$. The matching procedure mitigates the impact of this approximation, as it modifies one of the two sequences should they not match. Second, it is possible to start the MCMC directly from the hyperedge configuration of a real hypergraph, as we illustrate in our experiments with real data in \cref{sec:real data}. This effectively corresponds to fixing the sequences $\degseq, \dimseq$ to those of the data, which are necessarily consistent. These are then preserved during MCMC while the hyperedges are mixed. Hence, in this case there is no need for a matching algorithm. We thus assume that a user would like to start from a desired set of values for both quantities, without worrying about their consistency.

We now describe our proposed matching algorithm. If there exists at least one hypergraph with sequences $\degseq, \dimseq$, we call $\degseq$ and $\dimseq$ \emph{compatible}. Our algorithm guarantees that, if $\degseq, \dimseq$ are compatible, one hypergraph with such sequences will be produced. If they are not compatible, however \textemdash as is the case for most samples from \cref{sec supp: sampling step 1}\textemdash
only one of them can be preserved. For this reason, the user is required to specify which sequence needs to be preserved; we refer to this as the priority sequence. The algorithm dynamically modifies $\degseq, \dimseq$ until exhaustion of the priority sequence, if the sequences are compatible then no modification to the other sequence will be made, as they will terminate together. 
Intuitively, our algorithm works as follows. It extracts a hyperedge of a given size from the nodes with the highest available degrees, until exhaustion of the size sequence $\dimseq$. If $\dimseq$ is the priority sequence and no nodes are available, random nodes are extracted to satisfy the required hyperedge sizes, otherwise smaller hyperedges might be produced.
After exhaustion of the size sequence, if $\degseq$ is still not exhausted, but it is the priority sequence, keep drawing hyperedges to exhaust $\degseq$ (even if $\dimseq$ is not preserved). The function described in \cref{alg: extract hye fn} draws the hyperedges and prioritizes $\degseq$ or $\dimseq$ depending on the priority sequence. We present a pseudocode description of the matching algorithm in \cref{alg: sequence merging}.

\begin{algorithm}
\caption{Hyperedge construction \\from sequences}\label{alg: sequence merging}
\KwInput{Degree sequence $\degseq$,  size sequence $\dimseq$, priority sequence choice $pr \,\in \ccup{\text{degseq, dimseq}}$} 
\KwResult{List of hyperedges $L$} 
\null
\Comment{Iterate over hyperdge sizes and specified number of such hyperedges}
L = [ ] \\ 
\For{$s=2, \ldots, D$ 
}{
    \For{$j=1, \ldots, \dimseq_i$  
    }{
  	    hye = ExtractHye($s, \degseq, \dimseq, pr$) \\
  	    \If{len(hye) $>1$}{
  	        L $\gets$ L + [hye] \\
  	        }
  }  	
} 
\Comment{If the priority is the degree sequence, have all nodes reach the required degree}
\Comment{Call GeTwo the function counting the number of nodes with non-zero degrees in the sequence} 
\If{$pr=\text{degseq}$}{
    \Comment{While there are at least two nodes with non-zero remaining degree, produce hyperedges}
    \While{GeTwo($\degseq$) $\ge 2$}{
        maxdim $\gets \min$(GeTwo($\degseq$), D) \\
        $s \gets$ random\{2, \ldots, maxdim\} \\
        hye $\gets$ ExtractHye($s, \degseq, \dimseq, pr$) \\
        L $\gets$ L + [hye] \\
    }
}
\end{algorithm}

\begin{algorithm}
\caption{ExtractHye}\label{alg: extract hye fn}
\KwInput{Hyperedeg size $s$, degree sequence $\degseq$, size sequence $\dimseq$, priority sequence choice $pr \,\in \ccup{\text{degseq, dimseq}}$} 
\KwResult{A hyperedge hye. \\\hspace{14mm}The input $\degseq$ is modified in place.} 
\Comment{Extract one hyperedge while preserving the degree or size sequence} 
\If{$pr=\text{degseq}$}{
    hye $\gets$ \{$s$ \text{ nodes $v$ with the highest $\degseq_v$ values.} \\ \text{\hspace{12mm}If not enough nodes satisfy $\degseq_v>0$,} \text{\hspace{10mm}possibly return less than $s$ nodes}\}
} 
\ElseIf{$pr=\text{dimseq}$}{
    hye $\gets$ \{$s$ \text{ nodes $v$ with the highest $\degseq_v$ values.} \\ \text{\hspace{12mm}If not enough nodes satisfy $\degseq_v>0$,}\\\text{\hspace{12mm}select some random nodes with $\degseq_v^b=0$}\\\text{\hspace{12mm}untill $s$ nodes are chosen.}\}
}
\null
\Comment{Update the degree sequence by decreasing the degree of the selected nodes}
\For{v in hye}{
    \If{$\degseq_v > 0$}{
        $\degseq_v \gets \degseq_v - 1$
    }
}
\end{algorithm}

\section{Generation of the synthetic data}
\label{sec supp: generation of synthetic benchmark}
We include additional details for the generation of the data utilized in Section~\cref{sec: benchmarking community}. We generate data with $N=500$ nodes, $K=3$ equally-sized communities, and hard community assignments. The adjacency matrix $w$ is the $K \times K$ identity matrix.
Additionally, we condition on a size sequence, i.e. the count of hyperedges per hyperedge dimension, given by:
\texttt{
\{
2: 500,
3: 400,
4: 400,
5: 400,
6: 600,
7: 700,
8: 800,
9: 900,
10: 1000,
11: 1100,
12: 1200,
13: 1300,
14: 1400,
15: 1500
 \}
 }. 
 The expected degree resulting from such a sequence is $248.6$. Like for other experiments, we utilize the default MCMC configuration of $n_b=100000$ burn-in steps and $n_i=20000$ steps between samples.

\section{Matching sequences on the House Bills dataset}
\begin{figure}[t]
\hspace{-0.4cm}
\includegraphics[width=0.45\textwidth]{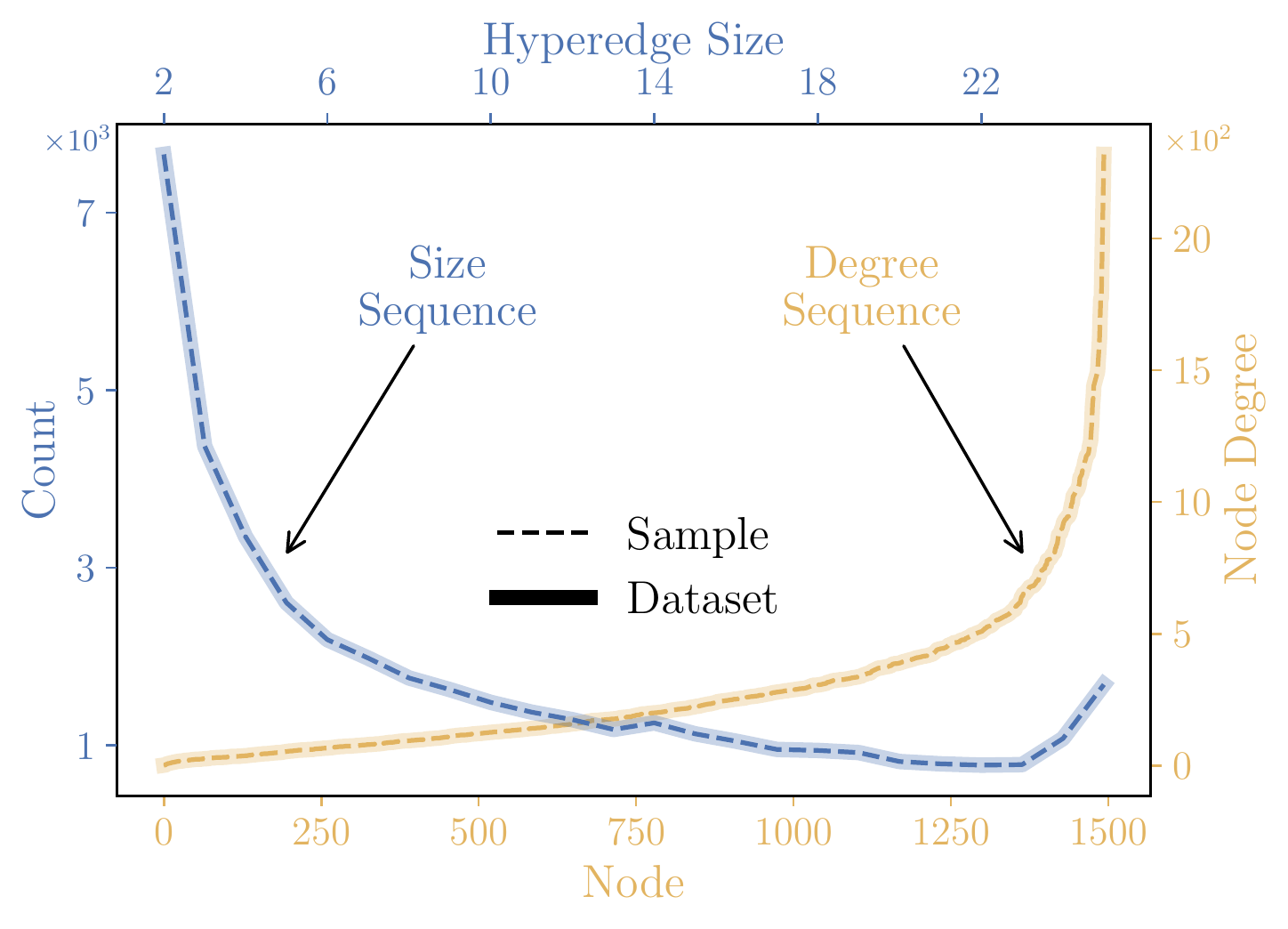}
\caption{
    \textbf{Correspondence of the degree and size sequences between data and samples}. We check for the correspondence between the sequences of the samples and the original House Bills dataset, which is utilized to initialize the MCMC procedure. Due to the properties of the reshuffling operator, the sequences need to coincide.
}
\label{fig: sequences house bills}
\end{figure}

In \cref{fig: sequences house bills} we show the perfect correspondence between the degree and size sequences as observed in the real data and in the samples. As we initialize our sampling procedure directly from the hyperedge configuration of the real data, such correspondence is guaranteed by the properties of the reshuffling operator. 

\section{Structural measures from configuration model}
\label{sec supp: experiments configuration model}

\begin{figure*}[t]
\hspace{-0.4cm}
\includegraphics[width=0.95\textwidth]{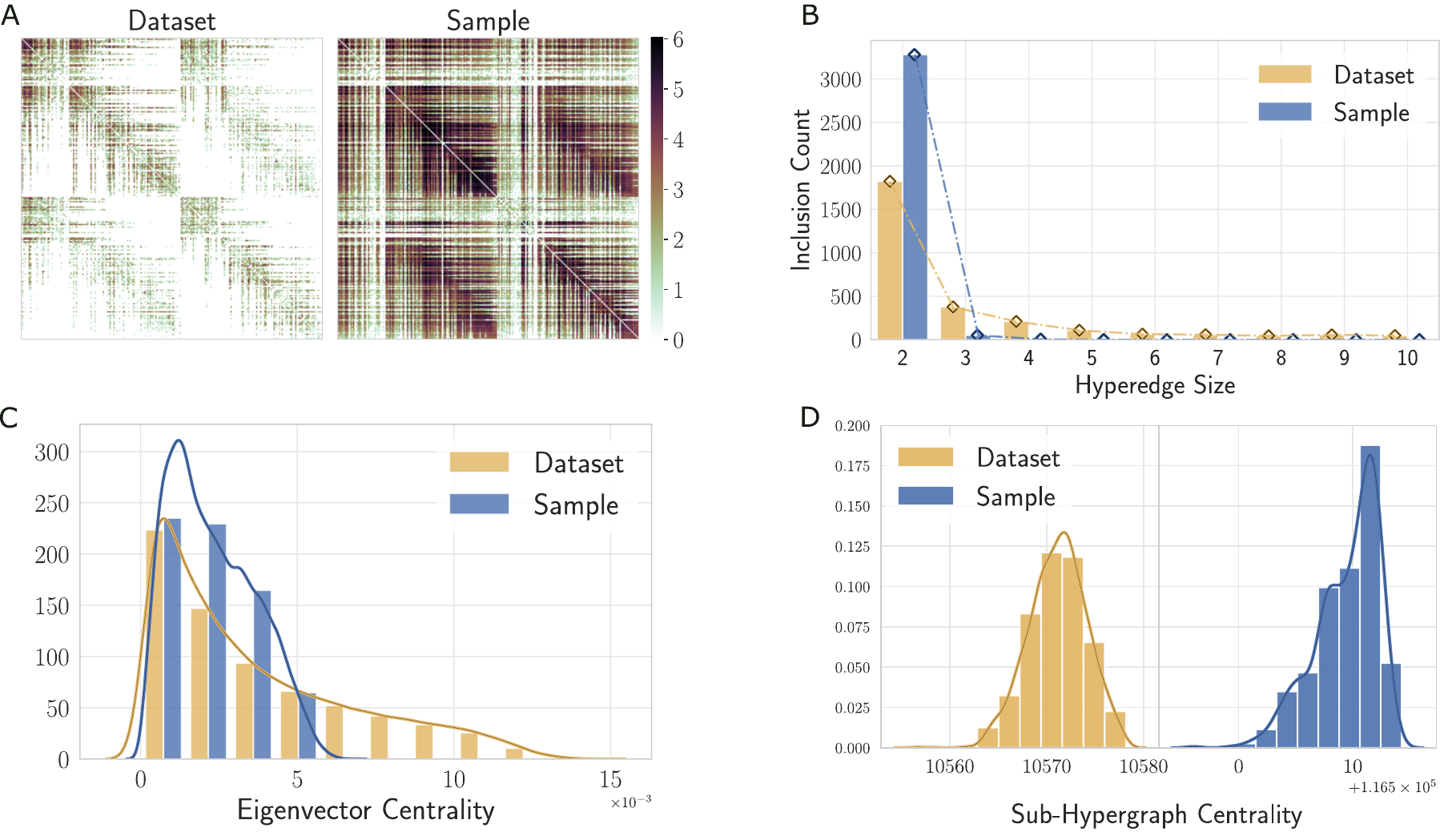}
\caption{
    \textbf{Comparing the statistics on real data and samples obtained from the configuration model}. We plot (A) the adjacency matrices, (B) the hyperedge inclusions occurences, (C) the hyperedge eigenvector centrality distribution and (D) the sub-hypergraph centrality distribution for the House Bills dataset. Here, samples are obtained via the hypergraph configuration model \cite{chodrow2020configuration}. Due to less structure being incorporated into the sampling procedure, samples and real data present substantial differences.
}
\label{fig: house bills configuration model}
\end{figure*}

    For comparison, here we run the experiments in \cref{sec: data vs sample stats}, but create samples via the hypergraph configuration model from Chodrow~\cite{chodrow2020configuration}, as opposed to our method. The configuration model takes a dataset and mixes its hyperedges preserving the initial degree and size sequences. In this sense, it can be seen as a less structured version of the method we propose here.
    In \cref{fig: house bills configuration model}, we show the adjacency matrix, hyperedge inclusions, hyperedge eigenvector centrality and sub-hypergraph centrality computed on samples based on the House Bills dataset. As can be observed, the samples obtained via the configuration model have less resemblance to the original dataset.

\bibliographystyle{ScienceAdvances}
\bibliography{bibliography}

\end{document}